\documentclass[showpacs,aps,graphicx,onecolumn,10pt]{revtex4-2}
\usepackage{amsmath}
\usepackage{mathrsfs}
\usepackage{amsfonts}
\usepackage{amssymb}
\usepackage{graphicx}
\usepackage{caption}
\usepackage{subfigure}
\usepackage{eufrak}
\usepackage{multirow}
\usepackage{float}
\usepackage[colorlinks,linkcolor=blue,anchorcolor=blue,citecolor=blue]{hyperref}
\usepackage{soul,color,xcolor}

\begin{document}

\title{Entanglement concentration of high-dimensional unknown partially entangled state}

\author{Si-Qi Du\textsuperscript{1}, Guo-Zhu Song\textsuperscript{2} and Hai-Rui Wei\textsuperscript{1,}}
\email[]{hrwei@ustb.edu.cn}
\address{\textsuperscript{\rm1} School of Mathematics and Physics, University of Science and Technology Beijing, Beijing 100083, China\\
\textsuperscript{\rm2} College of Physics and Materials Science, Tianjin Normal University, Tianjin 300387, China}

\begin{abstract}
High-dimensional quantum systems offer a number of advantages in larger information capacity, stronger noise resiliency, higher improved efficiency and accuracy over the qubit systems.
In quantum communication the maximally entangled states will inevitably become mixed states or less-entangled pure states by the channel noise during the practical transmission or storage.
We propose a universal scheme to concentrate nonlocal high-dimensional generalized Bell states with unknown parameters.
After the cross-Kerr nonlinearities, $X$-quadrature homodyne measurements, and single-partite projection measurements are performed only at Bob's site, a two-qutrit maximally entangled Bell state can be distilled, while previous entanglement concentration protocols (ECPs) mostly focused on two-level qubit systems.
The concentrated partially entangled qubit states, reserved as the by-product are the fascinating resources for some quantum information processing tasks.
Moreover, single-qutrit projection measurement, the key ingredient for our ECP with unknown parameters, are completed by using linear optical elements.
Additionally, linear optical high-dimensional ECP with known parameters are also designed.
\end{abstract}


\maketitle

\section{Introduction}
 
Quantum entanglement \cite{entanglement1,entanglement2,entanglement3,entanglement4} is a fundamental resource for many quantum information processing (QIP) tasks, such as measurement-based quantum computation \cite{one-way1,one-way2}, quantum teleportation \cite{teleportation1,teleportation2,teleportation3}, quantum dense coding \cite{coding1,coding4}, and quantum key distribution \cite{distribution1,distribution2,distribution3}.
Typically, maximally entangled states (such as $|\psi\rangle=1/\sqrt{m}\sum_{i=0}^{m-1}|i\rangle|i\rangle$) are often necessary in these tasks to ensure success, security, and efficiency.
However, in practical transmission and process, the maximally entangled state will be turned into less-entangled state or mixed state due to the inevitably noise channels \cite{noise1,noise3}.
Then, it will further lead to the insecurity and inefficiency of quantum communication schemes, the failure of quantum dense coding, and the lower fidelity of QIP tasks.
Fortunately, entanglement purification (EPP) \cite{EPP}, entanglement concentration (ECP) \cite{ECP}, entanglement distillation \cite{distillation1,distillation2}, and error correction \cite{correction1,correction2} have been recognized as powerful tools to remedy these  problems. 
Recently, the EPP with nonidentical copies of partially entangled state had a drastic progress both in theory and experiment \cite{EPP1,EPP2,EPP3}. 
Additionally, EPP and ECP are the important components of the quantum repeater.

ECP is first introduced by Bennett \emph{et al.} \cite{Bennett1996}, which is called the Schmidt projection protocol, to distill maximally entangled state from less-entangled pure ones.
Inspired by this pioneering work, a number of improved and interesting ECPs were proposed later with different physical systems, different entangled states, and different data buses \cite{dot2012,NV2014,dot2024,Du2024}.
It is shown that all the existing ECPs can be divided into two groups.
In the first group, the parameters (information) of the less-entangled state are accurately known by the long-distance parties beforehand, such as those in Refs. \cite{Zhao2001,Yamamoto2001,Deng2012,Sheng2012}.
In the other group, the parameters are unknown to all the parties, such as those in Refs. \cite{Yang2005,Sheng2008,Sheng2010,auxiliary2012}.
In general cases, the former can be achieved with a higher efficiency than that in the latter,  the copies of the nonmaximally state are required for that in the latter.
Recently, hyperentanglement concentration protocols (hyper-ECPs) \cite{Ren2013,Li2015} have attracted considerable attention because of its better security, larger channel capacity of quantum communication, higher power of quantum computation, and ability to efficiently solve certain QIP tasks that are intractable on a conventional parallel counterpart.
Nowadays, various interesting hyper-ECPs have been developed \cite{Li2016,Liu2023}.  
However, the hyper-ECPs are usually more intricate, as more than one degree of freedom (DOF) must be well operated independently and simultaneously.

Significant progress in parallel and hyper-parallel two-dimensional qubit systems have been made.
However, high-dimensional (multi-level or multi-state system) ECPs are rarely investigated \cite{Vidal1999,Matsumoto2007,Vaziri2003} even that high-dimensional quantum systems exhibit many remarkable advantages over qubit systems in performing some QIP tasks.
Compared to qubit systems, high-dimensional systems provide a larger Hilbert space to store and process information \cite{highdimensionals2}, and thus can further help to enhance noise resistance \cite{highnoise1,highnoise2}, improve efficient quantum simulation and computation \cite{simulation}, reduce circuit complexity, larger violation of Bell inequality \cite{Bell}, and simplify experimental setup \cite{highdimensional2,highdimensional3}.
Based on these peculiar features, high-dimensional systems benefit some applications. For example, high-dimensional quantum teleportation \cite{high-teleportation}, high-dimensional quantum state preparation \cite{high-generation1,high-generation2}, high-dimensional entangled state measurement \cite{HDBM1,HDBM2}, parallel teleportation of quantum operations \cite{Xiu2025}, high-dimensional quantum algorithm \cite{HDalgorithms}, and high-dimensional quantum gate \cite{HDgate1,HDgate2,HDgate3,HDgate4,HDgate5}.  
Early in 1999, Vidal \cite{Vidal1999} explored a high-dimensional ECP via Schmidt decomposition.
In 2003,  Vaziri \emph{et al.} \cite{Vaziri2003} experimental demonstrated a specific case of qutrit ECP. Different from above two ECPs with known information, later in 2007, Matsumoto \emph{et al.} \cite{Matsumoto2007} developed a universal distortion-free ECP without any information about the Schmidt basis.

In this paper, we proposed an ECP to distill the desired two-qutrit maximally entangled generalized Bell state from less-entangled state via cross-Kerr nonlinearities.
The qutrits (three-state or three-level system) are encoded in the spatial DOF of single photon systems.
Our qutrit ECP is universal due to the parameters of the partially entangled state are unknown to the two communication parties before ECP. 
All the operations introduced to complete the scheme are performed at only Bob's site.
Besides the desired maximally entangled states, the concentrated non-maximally entangled qubit states are the fascinating resource for some QIP tasks.
Then, the optical architecture is designed to implement single-partite projection measurement resorting to linear optical elements.
Finally, the high-dimensional ECP with known parameters case are constructed by using linear optics, and the feasibility of the above two schemes are evaluated.
In qubit ECP, a single copy of the bipartite entangled state is required to guarantee the desired maximally entangled state. In our qutrit ECP, two identical copies of the bipartite entangled states are required as the higher information capacity.
Compared with hyper-ECP, the operations in our scheme are relative ease to manipulate.
The Schmidt basis and Schmidt coefficients are known for Schmidt projection ECP \cite{Vidal1999,Vaziri2003}.  The high-dimensional ECP protocol in Ref. \cite{Matsumoto2007} is symmetry,  the action of the permutation and the complex multi-round local unitary transforms are required.

\section{ECP for two-photon two-qutrit unknown spatial entangled state}

In practical long-distance communication, the maximally entangled state $\frac{1}{\sqrt{3}}(|0_a 0_b\rangle+|1_a 1_b\rangle+|2_a 2_b\rangle)$ may be degraded to less-entangled state $|\psi\rangle$ due to the noisy channels.
Here
\begin{equation}\label{eq1}
    |\psi\rangle = \alpha|0_a 0_b\rangle+\beta|1_a 1_b\rangle+\gamma|2_a 2_b\rangle.
\end{equation}
Subscripts $a$ and $b$ label the single-photons $a$ and $b$, respectively.
$|0\rangle$, $|1\rangle$, and $|2\rangle$ denote the three spatial modes of the single-photons. 
The parameters $\alpha$, $\beta$, and $\gamma$ satisfy the normalization relation $|\alpha|^2+|\beta|^2+|\gamma|^2=1$.
Single-photons $a$ and $b$ held by Alice and Bob, respectively.

\begin{figure}[htbp]
\centering\includegraphics[width=12.5 cm]{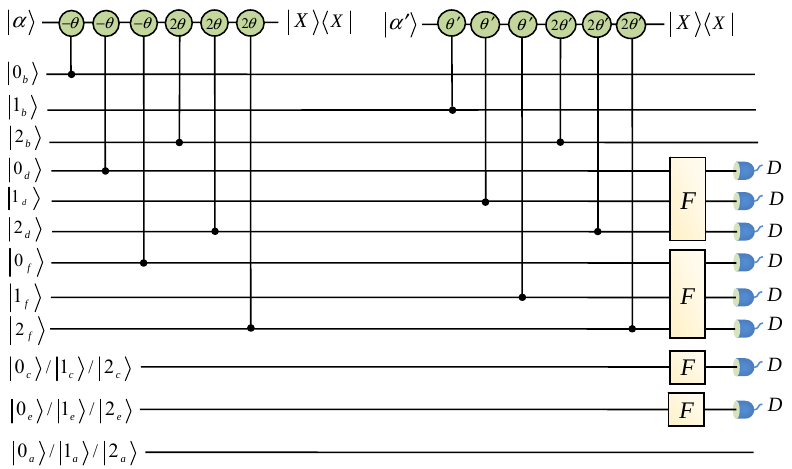}
\caption{A schematic diagram of the operations performed by Alice and Bob to concentrate a two-qutrit generalized state Bell via weak cross-Kerr nonlinearities. $F$ represents the single-qutrit spatial-based Fourier transformation. $D$ is single-photon detector.}
\label{Fig1}
\end{figure}

The procedure of our qutrit ECP assisted by two identical cross-Kerr nonlinearities is shown in Fig. \ref{Fig1}.
The Hamiltonian for describing the cross-Kerr nonlinearity is given by
\begin{equation}\label{eq2}
    H = \hbar\mathcal{X}{a_s}^\dagger a_s {a_p}^\dagger a_p.
\end{equation}
Here ${a_s}^\dagger$ (${a_p}^\dagger$) and $a_s$ ($a_p$) are the creation and annihilation operations for the signal (probe) state.
$\mathcal{X}$ is the coupling strength of the nonlinearity which is determined by the material. 
Kok \emph{et al.} \cite{Kok1,Kok2} showed that even the largest natural cross-Kerr nonlinearities are extremely weak $\mathcal{X}^{(3)}\approx10^{-22}$ $\text{m}^2 \text{V}^{-2}$, and
the Kerr phase shift is only $\tau\approx 10^{-18}$ in the optical single-photon regime with a mode volume of approximately 0.1 cm$^{3}$.

After the signal state $|n\rangle_s$ interacts with the coherent state $|\alpha\rangle$, the Kerr nonlinearity induces transformation
\begin{equation}\label{eq3}
    |n\rangle_s|\alpha\rangle_p \rightarrow |n\rangle_s|\alpha e^{\mathrm{i}n\theta}\rangle,
\end{equation}
where $|n\rangle_s$ represents the Fock state (photon number state) on the signal space mode, and $n$ is the number of photons in that mode.  $\theta=\mathcal{X}t$, and $t$ is the interaction time.  
In 2001, Lukin \emph{et al}. \cite{EIT1} pointed that the much large cross-Kerr nonlinearities of $\theta \approx 10^{-2}$ can be created by using electromagnetically induced transparency (EIT) materials. 
Recently, utilizing EIT technique, $\theta=18$ $\mu$rad are obtained by Feizpour \emph{et al}. \cite{EIT2}.

As shown in Fig. \ref{Fig1}, in order to distill the maximally entangled state $\frac{1}{\sqrt{3}}(|0_a 0_b\rangle+|1_a 1_b\rangle+|2_a 2_b\rangle)$ from a set of photon pairs in the state $|\psi\rangle$ described in Eq. \eqref{eq1}, two identical less-entangled photon pairs ($c$, $d$) and ($e$, $f$) should be introduced. Here ($c$, $d$) and ($e$, $f$) are considered as
\begin{equation}\label{eq4}
    |\psi'\rangle = \alpha|0_c 0_d\rangle+\beta|1_c 1_d\rangle+\gamma|2_c 2_d\rangle,
\end{equation}
\begin{equation}\label{eq5}
    |\psi''\rangle = \alpha|0_e 0_f\rangle+\beta|1_e 1_f\rangle+\gamma|2_e 2_f\rangle.
\end{equation}
Single-photons $c$ and $e$ belong to Alice,  $d$ and $f$ belong to Bob. 
The information of $|\psi'\rangle_{ab}$ and $|\psi''\rangle_{ab}$ all are unknown to Alice and Bob.

Firstly, Bob leads photons $b$, $d$, and $f$ to couple an optical mode in a coherent state $|\alpha\rangle$. In detail,
$|0_b\rangle$, $|0_d\rangle$, and $|0_f\rangle$ cause the coherent state $|\alpha\rangle$ to pick up a phase shift $-\theta$, respectively;
$|2_b\rangle$, $|2_d\rangle$, and $|2_f\rangle$  pick up a phase shift $2\theta$, respectively;
Others pick up no phase shift.
Hence, above interactions transform the state of the whole system composed of three photon pairs $(a,b)$, $(c,d)$, $(e,f)$ and the coherent state $|\alpha\rangle$ from the original state $|\Psi_0\rangle$ into $|\Psi_1\rangle$.
Here
\begin{equation}\label{eq6}
    |\Psi_0\rangle = |\psi\rangle \otimes |\psi'\rangle \otimes |\psi''\rangle \otimes |\alpha\rangle \otimes |\alpha'\rangle.
\end{equation}
\begin{equation}                         \label{eq7}
  \begin{split}
|\Psi_1\rangle =& \big(\alpha^3 |0_a 0_b 0_c 0_d 0_e 0_f\rangle |\alpha e^{-\mathrm{i}3\theta}\rangle + \beta^3 |1_a 1_b 1_c 1_d 1_e 1_f\rangle |\alpha\rangle 
                    + \gamma^3           |2_a 2_b 2_c 2_d 2_e 2_f\rangle |\alpha e^{\mathrm{i}6\theta}\rangle \\
               &+ \alpha^2\beta     \big(|0_a 0_b 1_c 1_d 0_e 0_f\rangle +  |0_a 0_b 0_c 0_d 1_e 1_f\rangle
                                      +  |1_a 1_b 0_c 0_d 0_e 0_f\rangle \big) |\alpha e^{-2\mathrm{i}\theta}\rangle \\
               &+ \alpha\beta^2     \big(|0_a 0_b 1_c 1_d 1_e 1_f\rangle +  |1_a 1_b 0_c 0_d 1_e 1_f\rangle
                                      +  |1_a 1_b 1_c 1_d 0_e 0_f\rangle \big) |\alpha e^{-\mathrm{i}\theta}\rangle\\
               &+ \alpha\beta\gamma \big(|0_a 0_b 1_c 1_d 2_e 2_f\rangle  + |0_a 0_b 2_c 2_d 1_e 1_f\rangle
                                      +  |1_a 1_b 0_c 0_d 2_e 2_f\rangle \\
               &+                        |1_a 1_b 2_c 2_d 0_e 0_f\rangle +  |2_a 2_b 0_c 0_d 1_e 1_f\rangle
                                      +  |2_a 2_b 1_c 1_d 0_e 0_f\rangle \big) |\alpha e^{\mathrm{i}\theta}\rangle \\
               &+ \alpha^2\gamma    \big(|0_a 0_b 2_c 2_d 0_e 0_f\rangle +  |2_a 2_b 0_c 0_d 0_e 0_f\rangle
                                      +  |0_a 0_b 0_c 0_d 2_e 2_f\rangle \big) |\alpha\rangle \\
               &+  \alpha\gamma^2   \big(|0_a 0_b 2_c 2_d 2_e 2_f\rangle+  |2_a 2_b 0_c 0_d 2_e 2_f\rangle
                                      +  |2_a 2_b 2_c 2_d 0_e 0_f\rangle \big) |\alpha e^{3\mathrm{i}\theta}\rangle  \\
               &+  \beta\gamma^2    \big(|1_a 1_b 2_c 2_d 2_e 2_f\rangle +  |2_a 2_b 1_c 1_d 2_e 2_f\rangle
                                      +  |2_a 2_b 2_c 2_d 1_e 1_f\rangle \big) |\alpha e^{4\mathrm{i}\theta}\rangle \\
               &+  \beta^2\gamma    \big(|1_a 1_b 1_c 1_d 2_e 2_f\rangle +  |1_a 1_b 2_c 2_d 1_e 1_f\rangle
                                      +  |2_a 2_b 1_c 1_d 1_e 1_f\rangle \big) |\alpha e^{2\mathrm{i}\theta}\rangle\big)|\alpha'\rangle.
  \end{split}
\end{equation}

Secondly, Bob leads $|1_b\rangle$, $|1_d\rangle$, $|1_f\rangle$, $|2_b\rangle$, $|2_d\rangle$, and $|2_f\rangle$ to another coherent state $|\alpha'\rangle$.
In detail,
$|1_b\rangle$, $|1_d\rangle$, and $|1_f\rangle$ pick up a phase shift $-\theta'$, respectively;
$|2_b\rangle$, $|2_d\rangle$, and $|2_f\rangle$ pick up a phase shift $2\theta'$, respectively.
And then, Eq. \eqref{eq7} is changed to
\begin{equation} \label{eq8}
 \begin{split}
|\Psi_2\rangle = \;& \alpha^2\beta  \big(|0_a 0_b 1_c 1_d 0_e 0_f\rangle +  |0_a 0_b 0_c 0_d 1_e 1_f\rangle
                                      +  |1_a 1_b 0_c 0_d 0_e 0_f\rangle \big) |\alpha e^{-2\mathrm{i}\theta}\rangle |\alpha'e^{\mathrm{i}\theta'}\rangle \\
               &+ \alpha\beta^2     \big(|0_a 0_b 1_c 1_d 1_e 1_f\rangle +  |1_a 1_b 0_c 0_d 1_e 1_f\rangle
                                      +  |1_a 1_b 1_c 1_d 0_e 0_f\rangle \big) |\alpha e^{-\mathrm{i}\theta}\rangle |\alpha'e^{\mathrm{i}2\theta'}\rangle \\
               &+ \alpha\beta\gamma \big(|0_a 0_b 1_c 1_d 2_e 2_f\rangle  + |0_a 0_b 2_c 2_d 1_e 1_f\rangle
                                      +  |1_a 1_b 0_c 0_d 2_e 2_f\rangle \\
               &+                        |1_a 1_b 2_c 2_d 0_e 0_f\rangle +  |2_a 2_b 0_c 0_d 1_e 1_f\rangle
                                      +  |2_a 2_b 1_c 1_d 0_e 0_f\rangle \big) |\alpha e^{\mathrm{i}\theta}\rangle |\alpha'e^{\mathrm{i}3\theta'}\rangle \\
               &+ \alpha^2\gamma    \big(|0_a 0_b 2_c 2_d 0_e 0_f\rangle +  |2_a 2_b 0_c 0_d 0_e 0_f\rangle
                                      +  |0_a 0_b 0_c 0_d 2_e 2_f\rangle \big) |\alpha\rangle |\alpha'e^{\mathrm{i}2\theta'}\rangle \\
               &+  \alpha\gamma^2   \big(|0_a 0_b 2_c 2_d 2_e 2_f\rangle+  |2_a 2_b 0_c 0_d 2_e 2_f\rangle
                                      +  |2_a 2_b 2_c 2_d 0_e 0_f\rangle \big) |\alpha e^{\mathrm{i}3\theta}\rangle |\alpha'e^{\mathrm{i}4\theta'}\rangle \\
               &+  \beta\gamma^2    \big(|1_a 1_b 2_c 2_d 2_e 2_f\rangle +  |2_a 2_b 1_c 1_d 2_e 2_f\rangle
                                      +  |2_a 2_b 2_c 2_d 1_e 1_f\rangle \big) |\alpha e^{\mathrm{i}4\theta}\rangle |\alpha'e^{\mathrm{i}5\theta'}\rangle \\
               &+  \beta^2\gamma    \big(|1_a 1_b 1_c 1_d 2_e 2_f\rangle +  |1_a 1_b 2_c 2_d 1_e 1_f\rangle
                                      +  |2_a 2_b 1_c 1_d 1_e 1_f\rangle \big) |\alpha e^{\mathrm{i}2\theta}\rangle |\alpha'e^{\mathrm{i}4\theta'}\rangle  \\
               &+  \alpha^3              |0_a 0_b 0_c 0_d 0_e 0_f\rangle |\alpha e^{-\mathrm{i}3\theta}\rangle |\alpha'\rangle + \beta^3 |1_a 1_b 1_c 1_d 1_e 1_f\rangle |\alpha\rangle |\alpha' e^{\mathrm{i}3\theta'}\rangle \\
               &+  \gamma^3              |2_a 2_b 2_c 2_d 2_e 2_f\rangle |\alpha e^{\mathrm{i}6\theta}\rangle |\alpha' e^{\mathrm{i}6\theta'}\rangle.
 \end{split}
\end{equation}

Thirdly, Bob performs two projection measurements ($X$-quadrature homodyne measurements) $|n\rangle\langle n|$ on the two output qutrit beams.
Based on Eq. \eqref{eq8}, one can see that the three photon pairs will collapse to the following different output states. More into the following details.

If the detection results are ($|\alpha e^{\mathrm{i}\theta}\rangle$, $|\alpha' e^{\mathrm{i}3\theta'}\rangle$), the state shown in Eq. \eqref{eq8} will collapse to the non-normalized state
\begin{equation}\label{eq9}
  \begin{split}
   |\Psi_{3_0}\rangle =\;& \alpha\beta\gamma  (|0_a 0_b 1_c 1_d 2_e 2_f\rangle
                                            +  |0_a 0_b 2_c 2_d 1_e 1_f\rangle
                                            +  |1_a 1_b 0_c 0_d 2_e 2_f\rangle  \\
                                           &+  |1_a 1_b 2_c 2_d 0_e 0_f\rangle
                                            +  |2_a 2_b 0_c 0_d 1_e 1_f\rangle
                                            +  |2_a 2_b 1_c 1_d 0_e 0_f\rangle).
 \end{split}
\end{equation}

If the detection results are ($|\alpha e^{-\mathrm{i}2\theta}\rangle$, $|\alpha' e^{\mathrm{i}\theta'}\rangle$), the state shown in Eq. \eqref{eq8} will collapse to the non-normalized state
\begin{equation}\label{eq10}
    |\Psi_{3_1}\rangle  = \alpha^2\beta  (|0_a 0_b 1_c 1_d 0_e 0_f\rangle
                                        + |0_a 0_b 0_c 0_d 1_e 1_f\rangle
                                        + |1_a 1_b 0_c 0_d 0_e 0_f\rangle).
\end{equation}

If the detection results are ($|\alpha e^{-\mathrm{i}\theta}\rangle$, $|\alpha' e^{\mathrm{i}2\theta'}\rangle$), the state shown in Eq. \eqref{eq8} will collapse to the non-normalized state
\begin{equation}\label{eq11}
    |\Psi_{3_2}\rangle  = \alpha\beta^2  (|0_a 0_b 1_c 1_d 1_e 1_f\rangle
                                        + |1_a 1_b 0_c 0_d 1_e 1_f\rangle
                                        + |1_a 1_b 1_c 1_d 0_e 0_f\rangle).
\end{equation}

If the detection results are ($|\alpha e^{\mathrm{i}\theta}\rangle$, $|\alpha' e^{\mathrm{i}2\theta'}\rangle$), the state shown in Eq. \eqref{eq8} will collapse to the non-normalized state
\begin{equation}\label{eq12}
    |\Psi_{3_3}\rangle  =  \alpha^2\gamma (|0_a 0_b 2_c 2_d 0_e 0_f\rangle
                                         + |2_a 2_b 0_c 0_d 0_e 0_f\rangle
                                         + |0_a 0_b 0_c 0_d 2_e 2_f\rangle).
\end{equation}

If the detection results are ($|\alpha e^{\mathrm{i}3\theta}\rangle$, $|\alpha' e^{\mathrm{i}4\theta'}\rangle$), the state shown in Eq. \eqref{eq8}  will collapse to the non-normalized state
\begin{equation}\label{eq13}
    |\Psi_{3_4}\rangle  =  \alpha\gamma^2  (|0_a 0_b 2_c 2_d 2_e 2_f\rangle
                                          + |2_a 2_b 0_c 0_d 2_e 2_f\rangle
                                          + |2_a 2_b 2_c 2_d 0_e 0_f\rangle).
\end{equation}

If the detection results are ($|\alpha e^{\mathrm{i}4\theta}\rangle$, $|\alpha' e^{\mathrm{i}5\theta'}\rangle$), the state shown in Eq. \eqref{eq8} will collapse to the non-normalized state
\begin{equation} \label{eq14}
    |\Psi_{3_5}\rangle  =  \beta\gamma^2  (|1_a 1_b 2_c 2_d 2_e 2_f\rangle
                                         + |2_a 2_b 1_c 1_d 2_e 2_f\rangle
                                         + |2_a 2_b 2_c 2_d 1_e 1_f\rangle).
\end{equation}

If the detection results are ($|\alpha e^{\mathrm{i}2\theta}\rangle$, $|\alpha' e^{\mathrm{i}4\theta'}\rangle$), the state shown in Eq. \eqref{eq8} will collapse to the non-normalized state
\begin{equation}\label{eq15}
    |\Psi_{3_6}\rangle  =  \beta^2\gamma  (|1_a 1_b 1_c 1_d 2_e 2_f\rangle
                                         + |1_a 1_b 2_c 2_d 1_e 1_f\rangle
                                         + |2_a 2_b 1_c 1_d 1_e 1_f\rangle).
\end{equation}

If the detection results are ($|\alpha e^{-\mathrm{i}3\theta}\rangle$, $|\alpha'\rangle$), the state shown in Eq. \eqref{eq8} will collapse to the non-normalized state
\begin{equation}\label{eq16}
    |\Psi_{3_7}\rangle = \alpha^3 |0_a 0_b 0_c 0_d 0_e 0_f\rangle.
\end{equation}

If the detection results are ($|\alpha\rangle$, $|\alpha' e^{\mathrm{i}3\theta'}\rangle$), the state shown in Eq. \eqref{eq8} will collapse to the non-normalized state
\begin{equation}\label{eq17}
    |\Psi_{3_8}\rangle = \beta^3 |1_a 1_b 1_c 1_d 1_e 1_f\rangle.
\end{equation}

If the detection results are ($|\alpha e^{\mathrm{i}6\theta}\rangle$,  $|\alpha' e^{\mathrm{i}6\theta'}\rangle$), the state shown in Eq. \eqref{eq8} will collapse to the non-normalized state
\begin{equation}\label{eq18}
    |\Psi_{3_9}\rangle = \gamma^3|2_a 2_b 2_c 2_d 2_e 2_f\rangle.
\end{equation}

Fourthly, two communication parities Alice and Bob measure photon pairs ($c$, $e$) and ($d$, $f$) in the orthogonal basis $\{|\phi_0\rangle, |\phi_1\rangle, |\phi_2\rangle\}$. Here
\begin{equation}\label{eq19}
  \begin{split}
     &|\phi_0\rangle = \frac{1}{\sqrt{3}}(|0\rangle + |1\rangle + |2\rangle), \\
     &|\phi_1\rangle = \frac{1}{\sqrt{3}}(|0\rangle + e^{\frac{\mathrm{i}2\pi}{3}} |1\rangle
                                                    + e^{\frac{\mathrm{i}4\pi}{3}} |2\rangle), \\
     &|\phi_2\rangle = \frac{1}{\sqrt{3}}(|0\rangle + e^{\frac{\mathrm{i}4\pi}{3}} |1\rangle
                                                    + e^{\frac{\mathrm{i}2\pi}{3}} |2\rangle).
  \end{split}
\end{equation}

Based on above measurement results, Alice or Bob performs some classical feed-forward operations on photons $a$ and $b$ to complete our ECP.
And then, $|\Psi_{3_0}\rangle$-$|\Psi_{3_9}\rangle$ are converted to the following normalization state (see Table \ref{Table1} for more detail)
\begin{equation}\label{eq20}
    |\Psi_{3_0}\rangle\rightarrow |\Psi_{4_0}\rangle = \frac{1}{\sqrt{3}} (|0_a 0_b\rangle + |1_a 1_b\rangle + |2_a 2_b\rangle),
\end{equation}
which take place with a probability of $P_{4_0}=6|\alpha\beta\gamma|^{2}$.
\begin{equation}\label{eq21}
    |\Psi_{3_1}\rangle \rightarrow |\Psi_{4_1}\rangle = \frac{2}{\sqrt{5}} |0_a 0_b\rangle + \frac{1}{\sqrt{5}} |1_a 1_b\rangle,
\end{equation}
which take place with a success probability of $P_{4_1}= 3|\alpha^2\beta|^2$.
\begin{equation}\label{eq22}
    |\Psi_{3_2}\rangle \rightarrow |\Psi_{4_2}\rangle = \frac{1}{\sqrt{5}} |0_a 0_b\rangle + \frac{2}{\sqrt{5}} |1_a 1_b\rangle,
\end{equation}
which take place with a success probability of $P_{4_2}= 3|\alpha\beta^2|^2$.
\begin{equation}\label{eq23}
    |\Psi_{3_3}\rangle \rightarrow |\Psi_{4_3}\rangle = \frac{2}{\sqrt{5}} |0_a 0_b\rangle + \frac{1}{\sqrt{5}} |2_a 2_b\rangle,
\end{equation}
which take place with a success probability of $P_{4_3}= 3|\alpha^2\gamma|^2$.
\begin{equation}\label{eq24}
    |\Psi_{3_4}\rangle \rightarrow |\Psi_{4_4}\rangle = \frac{1}{\sqrt{5}} |0_a 0_b\rangle + \frac{2}{\sqrt{5}} |2_a 2_b\rangle,
\end{equation}
which take place with a success probability of $P_{4_4}= 3|\alpha\gamma^2|^2$.
\begin{equation}\label{eq25}
    |\Psi_{3_5}\rangle \rightarrow |\Psi_{4_5}\rangle = \frac{1}{\sqrt{5}} |1_a 1_b\rangle + \frac{2}{\sqrt{5}} |2_a 2_b\rangle,
\end{equation}
which take place with a success probability of $P_{4_5}= 3|\beta\gamma^2|^2$.
\begin{equation}\label{eq26}
    |\Psi_{3_6}\rangle \rightarrow |\Psi_{4_6}\rangle = \frac{2}{\sqrt{5}} |1_a 1_b\rangle + \frac{1}{\sqrt{5}} |2_a 2_b\rangle,
\end{equation}
which take place with a success probability of $P_{4_6}= 3|\beta^2\gamma|^2$.
\begin{equation}\label{eq27}
    |\Psi_{3_7}\rangle \rightarrow |\Psi_{4_7}\rangle=\alpha^3 |0_a 0_b\rangle,
\end{equation}
which take place with a success probability of $P_{4_7}=|\alpha^3|^2$.
\begin{equation}\label{eq28}
    |\Psi_{3_8}\rangle \rightarrow |\Psi_{4_8}\rangle=\beta^3 |1_a 1_b \rangle.
\end{equation}
which take place with a success probability of $P_{4_8}=|\beta^3|^2$.
\begin{equation}\label{eq29}
    |\Psi_{3_9}\rangle \rightarrow |\Psi_{4_9}\rangle=\gamma^3|2_a 2_b \rangle.
\end{equation}
which take place with a success probability of $P_{4_9}=|\gamma^3|^2$.

\smallskip

\begin{table}[htbp]
  \caption{The correspondences between coefficients and phase shifts in the proposed ECP.}\label{Table1}
  \centering
\begin{tabular}{ccccccccccc}
\hline\hline
   Coefficient & $\alpha^3$ & $\alpha^2\beta$ & $\alpha\beta^2$ & $\alpha\beta\gamma$ & $\alpha^2\gamma$ & $\alpha\gamma^2$ & $\beta\gamma^2$ & $\beta^2\gamma$ & $\beta^3$ & $\gamma^3$\\
\hline
 First phase shift   & $-3\theta$                   & $-2\theta$       & $-\theta$          & $\theta$          & $0$              & $3\theta$       & $4\theta$        & $2\theta$       & 
      $0$            & $6\theta$   \\

 Second phase shift  & $0$                          & $\theta'$        & $2\theta'$         & $3\theta'$        & $2\theta'$       & $4\theta'$      & $5\theta'$       & $4\theta'$      &
    $3\theta'$       & $6\theta'$  \\

 Output states       & $\Psi_{4_7}$                 & $\Psi_{4_1}$     & $\Psi_{4_2}$       & $\Psi_{4_0}$      & $\Psi_{4_3}$     & $\Psi_{4_4}$    & $\Psi_{4_5}$     & $\Psi_{4_6}$    &
    $\Psi_{4_8}$     & $\Psi_{4_9}$  \\

 Success probability & $P_{4_7}$                    & $P_{4_1}$        & $P_{4_2}$          & $P_{4_0}$         & $P_{4_3}$        & $P_{4_4}$       & $P_{4_5}$        & $P_{4_6}$       & 
      $P_{4_8}$      & $P_{4_9}$\\
\hline\hline
\end{tabular}
\end{table}

To clarify our description, we will take $|\Psi_{3_0}\rangle$ as an example.
In the basis $\{|\phi_0\rangle, |\phi_1\rangle, |\phi_2\rangle\}$, state $|\Psi_{3_0}\rangle$ can be written as
\begin{equation}\label{eq30}
  \begin{split}
  |\Psi_{3_0}\rangle=\;& \frac{1}{\sqrt{3}}|\phi_k\rangle_c|\phi_l\rangle_e|\phi _m\rangle_d|\phi_n\rangle_f
                                       \big((e^{\frac{\mathrm{i}(l\oplus n)2\pi}{3}} |2_a 2_b\rangle
                                           + e^{\frac{\mathrm{i}(l\oplus n)4\pi}{3}} |1_a 1_b\rangle) \\
                     & +  (|2_a 2_b\rangle + e^{\frac{\mathrm{i}(l\oplus n)4\pi}{3}} |0_a 0_b\rangle) e^{\frac{\mathrm{i}(k\oplus m)2\pi}{3}} 
                       +  (|1_a 1_b\rangle + e^{\frac{\mathrm{i}(l\oplus n)2\pi}{3}} |0_a 0_b\rangle) e^{\frac{\mathrm{i}(k\oplus m)4\pi}{3}}\big),
  \end{split}
\end{equation}
where $k,l,m,n \in(0,1,2)$.

Based on Eq. \eqref{eq30}, one can see that after the single-partite projection measurements are performed on photons $c$, $d$, $e$, and $f$, and some classical feed-forward operations are performed on photons $a$ and $b$, the two-qutrit maximally entangled state $|\Psi_{4_0}\rangle=\frac{1}{\sqrt{3}}(|0_a 0_b\rangle + |1_a 1_b\rangle + |2_a 2_b\rangle)$ can be obtained, see Table \ref{Table2} and Table \ref{Table3} for more detail. For example,
the scenario ($|\phi_1\rangle_d$, $|\phi_2\rangle_f$, $|\phi_2\rangle_c$, $|\phi_2\rangle_e$) means $|\Psi_{3_0}\rangle$ shown in Eq. \eqref{eq30} collapses to
\begin{equation}\label{eq31}
  |\Psi_{4_0}\rangle = \frac{1}{\sqrt{3}}(|0_a 0_b\rangle + |1_a 1_b\rangle + |2_a 2_b\rangle),
\end{equation}
with a probability of $2|\alpha\beta\gamma|^2$.

The scenario ($|\phi_1\rangle_d$, $|\phi_2\rangle_f$, $|\phi_0\rangle_c$, $|\phi_2\rangle_e$) means $|\Psi_{3_0}\rangle$ shown in Eq. \eqref{eq30} collapses to
\begin{equation}\label{eq32}
  |\Psi_{3_0}^{''}\rangle = \frac{1}{\sqrt{3}}(|0_a 0_b\rangle + e^{\frac{\mathrm{i}4\pi}{3}} |1_a 1_b\rangle + e^{\frac{\mathrm{i}2\pi}{3}} |2_a 2_b\rangle),
\end{equation}
with a probability of $\frac{2}{3}|\alpha\beta\gamma|^2$.
Obviously, $|\Psi_{3_0}^{''}\rangle$ can be converted into $|\Psi_{4_0}\rangle$ by applying single-qutrit operation $U_1$ given in Eq. \eqref{eq36} on photon $a$.

The scenario ($|\phi_1\rangle_d$, $|\phi_2\rangle_f$, $|\phi_1\rangle_c$, $|\phi_0\rangle_e$) means $|\Psi_{3_0}\rangle$ shown in Eq. \eqref{eq30} collapses to
\begin{equation}\label{eq33}
  |\Psi_{3_0}^{'''}\rangle  = \frac{1}{\sqrt{3}}(|0_a 0_b\rangle + e^{\frac{\mathrm{i}2\pi}{3}} |1_a 1_b\rangle + e^{\frac{\mathrm{i}4\pi}{3}} |2_a 2_b\rangle),
\end{equation}
with a probability of $\frac{2}{3}|\alpha\beta\gamma|^2$.
Then $|\Psi_{3_0}^{'''}\rangle$ can also be converted into $|\Psi_{4_0}\rangle$ by applying single-qutrit operation $U_2$ shown in Eq. \eqref{eq36} on photon $a$.

The scenario ($|\phi_2\rangle_d$, $|\phi_0\rangle_f$, $|\phi_1\rangle_c$, $|\phi_1\rangle_e$) means $|\Psi_{3_0}\rangle$ shown in Eq. \eqref{eq30} collapses to
\begin{equation}\label{eq34}
  |\Psi_{3_0}^{''''}\rangle  = \frac{1}{\sqrt{3}}(-|0_a 0_b\rangle + e^{\frac{\mathrm{i}5\pi}{3}} |1_a 1_b\rangle + e^{\frac{\mathrm{i}\pi}{3}} |2_a 2_b\rangle),
\end{equation}
with a probability of $\frac{4}{3}|\alpha\beta\gamma|^2$.
Then $|\Psi_{3_0}^{''''}\rangle$ can also be converted into $|\Psi_{4_0}\rangle$ by applying single-qutrit operation $U_3$ shown in Eq. \eqref{eq36} on photon $a$.

The scenario ($|\phi_2\rangle_d$, $|\phi_1\rangle_f$, $|\phi_1\rangle_c$, $|\phi_1\rangle_e$) means $|\Psi_{3_0}\rangle$ shown in Eq. \eqref{eq30} collapses to
\begin{equation}\label{eq35}
  |\Psi_{3_0}^{'''''}\rangle  = \frac{1}{\sqrt{3}}(-|0_a 0_b\rangle + e^{\frac{\mathrm{i}\pi}{3}} |1_a 1_b\rangle + e^{\frac{\mathrm{i}5\pi}{3}} |2_a 2_b\rangle),
\end{equation}
with a probability of $\frac{4}{3}|\alpha\beta\gamma|^2$.
Then $|\Psi_{3_0}^{'''''}\rangle$ can also be converted into $|\Psi_{4_0}\rangle$ by applying single-qutrit operation $U_4$ shown in Eq. \eqref{eq36} on photon $a$.

The single-qutrit spatial operations $U_0=I$, $U_1$, $U_2$, $U_3$, and $U_4$ are given by
\begin{equation}\label{eq36}
 \begin{split}
&U_1 = \left(
	\begin{array}{ccc}
    1  &  0                             &  0 \\
	0  &  e^{\frac{\mathrm{i}2\pi}{3}}  &  0 \\
	0  &  0                             &  e^{\frac{\mathrm{i}4\pi}{3}}
	\end{array}\right),\quad
U_2 = \left(
	\begin{array}{ccc}
    1  &  0                             &  0 \\
	0  &  e^{\frac{\mathrm{i}4\pi}{3}}  &  0 \\
	0  &  0                             &  e^{\frac{\mathrm{i}2\pi}{3}}
	\end{array}\right),\\
&U_3 = \left(
	\begin{array}{ccc}
    -1  &  0                             &  0 \\
	0   &  e^{\frac{\mathrm{i}\pi}{3}}   &  0 \\
	0   &  0                             &  e^{\frac{\mathrm{i}5\pi}{3}}
	\end{array}\right),\quad
U_4 = \left(
	\begin{array}{ccc}
    -1  &  0                             &  0 \\
	0   &  e^{\frac{\mathrm{i}5\pi}{3}}  &  0 \\
	0   &  0                             &  e^{\frac{\mathrm{i}\pi}{3}}
	\end{array}\right).
 \end{split}
\end{equation}

Putting Eq. \eqref{eq31}-\eqref{eq35} together, we can see that our qutrit ECP is completed with a probability of $6|\alpha\beta\gamma|^2$.

It is worthy to point out here that the states $|\Psi_{4_1}\rangle$-$|\Psi_{4_6}\rangle$ given in Eq. \eqref{eq21}- \eqref{eq26} all are acting on two-state quantum systems due to the two-state (i.e., $\{|0\rangle,|1\rangle\}$, $\{|0\rangle,|2\rangle\}$, or $\{|1\rangle,|2\rangle\}$) subsystems. 
These two-qubit non-maximally entangled states with known coefficients can become maximally entangled states by using split-based ECP \cite{Ren2013}.
They are also the fascinating resource for some qubit QIP tasks, such as circuit cutting \cite{cut}, deterministic hierarchical sharing \cite{DHS}, controller-independent quantum bidirectional communication \cite{CIQBC}, and quantum key distribution \cite{QKD}. 
The correspondences between the coefficients, $|X\rangle \langle X|$, the single-partite projection measurements, and the partially entangled states in qubit subsystem are shown in Table \ref{Table4}.

For scenario of the qubit ECP protocol with unknown parameters via cross-Kerr nonlinearity \cite{Sheng2008}, two identical input states are required. 
However, based on Eq. \eqref{eq1}, Eq. \eqref{eq4} and Eq. \eqref{eq5}, we can see that three identical input states are necessary for the qutrit ECP in the best case. 
The success probability of the qubit EPP can be further improved by iterating the entanglement concentration process \cite{Sheng2008}. 
Unfortunately, based on Eq. \eqref{eq20}-\eqref{eq29}, one can see that the success probability of our qutrit ECP can not be further improved by iterating the entanglement concentration process. 
That because no out coming photon pairs kept are in the state $ (\alpha^3|0_a 0_b 0_c 0_d 0_e 0_f\rangle+\beta^3|1_a 1_b 1_c 1_d 1_e 1_f\rangle+\gamma^3 |2_a 2_b 2_c 2_d 2_e 2_f\rangle)$.

\section{Implementation of three-dimensional Fourier Transformation with linear optical elements} \label{sec3}

One key element of our scheme is the projection measurements of photons $c$, $d$, $e$, and $f$ in the three-dimensional orthogonal basis $\{|\phi_0\rangle,|\phi_1\rangle,|\phi_2\rangle\}$. Fortunately, this task can be completed by employing spatial-based single-qutrit Fourier $U_F$ and single-photon detectors.
That is
\begin{equation}\label{37}
      \left(
        \begin{array}{c}
          |\phi_0\rangle \\
          |\phi_1\rangle \\
          |\phi_2\rangle \\
        \end{array}
      \right)
=
      U_F \cdot
      \left(
           \begin{array}{c}
             |0\rangle \\
             |1\rangle \\
             |2\rangle \\
           \end{array}
         \right)
=
		\frac{1}{\sqrt{3}}
     \left(
    \begin{array}{ccc}
       1 & 1                             & 1                            \\
       1 & e^{\frac{\mathrm{i}2\pi}{3}}  & e^{\frac{\mathrm{i}4\pi}{3}} \\
       1 & e^{\frac{\mathrm{i}4\pi}{3}}  & e^{\frac{\mathrm{i}2\pi}{3}} \\
    \end{array}
      \right)
\cdot
      \left(
           \begin{array}{c}
             |0\rangle \\
             |1\rangle \\
             |2\rangle \\
           \end{array}
         \right).
\end{equation}

\begin{figure}[htbp]
\centering\includegraphics[width=8.0 cm]{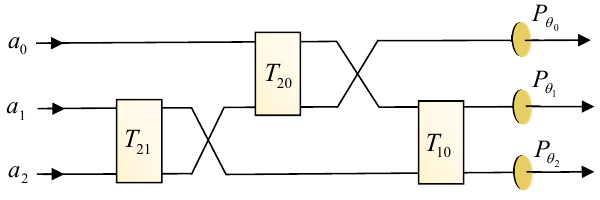}
\caption{A schematic diagram of the unitary transformation $U_F$ described in Eq. \eqref{37}. $T_{21}$, $T_{20}$ and $T_{10}$ are three variable beam splitters. $P_{\theta_0}=e^{\frac{\mathrm{i}\pi}{2}}$, $P_{\theta_1}=e^{\frac{\mathrm{i}4\pi}{3}}$ and $P_{\theta_2}=e^{\frac{\mathrm{i}2\pi}{3}}$ are phase shifters.}
\label{Fig2}
\end{figure}

Fig. \ref{Fig2} shows the setup for implementing the spatial-based unitary operation $U_F$,
which consists of three variable beam splitters $T_{21}$, $T_{20}$, $T_{10}$ and three phase shifters
$P_{\theta_0}=e^{\frac{\mathrm{i}\pi}{2}}$, $P_{\theta_1}=e^{\frac{\mathrm{i}4\pi}{3}}$, $P_{\theta_2}=e^{\frac{\mathrm{i}2\pi}{3}}$.
In the spatial-based basis $\{|0\rangle, |1\rangle, |2\rangle\}$, the unitary matrix of above $T_{21}$, $T_{20}$ and $T_{10}$ are given by
\begin{equation}\label{38}
 \begin{split}
  &T_{21}= \left(
  \begin{array}{ccc}
  1 & 0                                          & 0\\
  0 & e^{\mathrm{i}\varphi_{21}}\sin\omega_{21}  & \cos\omega_{21}\\
  0 & e^{\mathrm{i}\varphi_{21}}\cos\omega_{21}  & -\sin\omega_{21}
  \end{array}\right),\quad
T_{20}= \left(
  \begin{array}{ccc}
  e^{\mathrm{i}\varphi_{20}}\sin\omega_{20}  & 0 &  \cos\omega_{20}\\
              0                              & 1 &   0 \\
  e^{\mathrm{i}\varphi_{20}}\cos\omega_{20}  & 0 & -\sin\omega_{20}
  \end{array}\right),\\&
T_{10}= \left(
  \begin{array}{ccc}
  e^{\mathrm{i}\varphi_{10}}\sin\omega_{10} & \cos\omega_{10}  & 0\\
  e^{\mathrm{i}\varphi_{10}}\cos\omega_{10} & -\sin\omega_{10} & 0\\
  0                                         & 0                & 1
  \end{array}\right).
 \end{split}
\end{equation}

\begin{figure}[htbp]
\centering\includegraphics[width=9.0 cm]{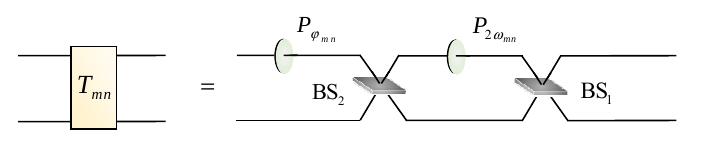}
\caption{Schematic diagram of the variable beam splitter $T_{mn}$ shown in Fig. \ref{Fig2}. BS is 50:50 beam splitter.  $P_{\varphi_{mn}}$ and $P_{2\omega_{mn}}$ can be achieved by employing phase shitters.}
\label{Fig3}
\end{figure}

\begin{figure}[htbp]
\centering\includegraphics[width=9.0 cm]{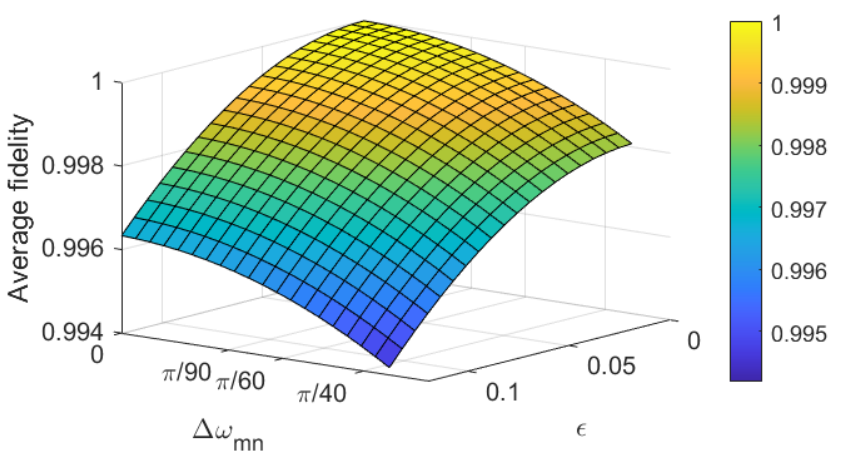}
\caption{Average fidelity of the block $T_{mn}$. Here $\Delta\omega_{mn}=\Delta\varphi_{mn}$, $\omega_{mn}=\frac{\pi}{4}$, and $\varphi_{mn}=\frac{\pi}{3}$ are taken.}
\label{Fig4}
\end{figure}

Inspired by Refs. \cite{UBS1,UBS2}, we design the quantum circuit for implementing $T_{21}$, $T_{20}$ and $T_{10}$, see Fig. \ref{Fig3}.
The parameter pairs
              ($\varphi_{21}=\frac{\pi}{3} $, $\omega_{21}=\frac{\pi}{4}$)   are taken for $T_{21}$;
              ($\varphi_{20}=\frac{2\pi}{3}$, $\omega_{20}=\arctan\sqrt{2}$) are taken for $T_{20}$;
              ($\varphi_{10}=\frac{3\pi}{2}$, $\omega_{10}=\frac{\pi}{4}$)   are taken for $T_{10}$. 
BS is 50:50 beam splitter.
The matrix representation of the BS and phase shifter are given by
\begin{equation}\label{eq39}
U_{\text{BS}} = \frac{1}{\sqrt{2}}
    \left(
      \begin{array}{cc}
        1              &    \mathrm{i} \\
        \text{i}       &        1 \\
      \end{array}
    \right), \quad
U_{P_{2\omega}} =
    \left(
      \begin{array}{cc}
        e^{\mathrm{i}2\omega}        &     0 \\
             0                       &     1 \\
      \end{array}
    \right), \quad
U_{P_\varphi} =
    \left(
      \begin{array}{cc}
        e^{\mathrm{i}\varphi}        &     0 \\
             0                       &     1 \\
      \end{array}
    \right).
\end{equation}
Unfortunately, in practical application, if the transmission rate of balanced BS deviates slightly from 50\% is $\epsilon$, then the matrix form of BS becomes \cite{imperfection}
\begin{equation}\label{eq40}
\widetilde{U}_{\text{BS}}=\frac{1}{\sqrt{{\epsilon}^{2}+2\epsilon+2}}
        \left(
           \begin{array}{cc}
            1 + \epsilon           &  \mathrm{i}   \\
	           \mathrm{i}          &  1+\epsilon \\
          \end{array}
       \right).
\end{equation}
If the phases difference between the two arms is $(\omega_{mn}-\Delta\omega_{mn})$ and $(\varphi_{mn}-\Delta\varphi_{mn})$, then the matrix form of two phase shifters become \cite{imperfection}
\begin{equation}\label{eq41}
\widetilde{U}_{P_{2\omega_{mn}}}=
        \left(
           \begin{array}{cc}
            e^{\mathrm{i}2(\omega_{mn}-\Delta\omega_{mn})}  &  0  \\
	                      0                                &  1  \\
          \end{array}
       \right),\quad
\widetilde{U}_{P_{\varphi_{mn}}}=
        \left(
           \begin{array}{cc}
            e^{\mathrm{i}(\varphi_{mn}-\Delta\varphi_{mn})}   &  0  \\
	                      0                                   &  1  \\
          \end{array}
       \right).
\end{equation}
The imperfection BSs and phase shifters will degrade the fidelity of the scheme. Without loss of generality, we consider the input state of Fig. \ref{Fig3} is
\begin{equation}\label{eq42}
|\zeta_{\text{in}}\rangle = \cos \alpha |0\rangle + \sin \alpha|1\rangle.
\end{equation}
The ideal output state will be
\begin{equation}\label{eq43}
\begin{split}
|\zeta_{\text{out}}\rangle_{\text{idea}} = & 
    \big(\frac{1}{2} ((e^{\mathrm{i}2\omega_{mn}} - 1) e^{\mathrm{i}\varphi_{mn}} \cos\alpha         + \mathrm{i}(e^{\text{i}2\omega_{mn}} + 1) \sin\alpha) \big)|0\rangle \\&
  + \big(\frac{1}{2} (\text{i}(e^{\mathrm{i}2\omega_{mn}} + 1)e^{\mathrm{i}\varphi_{mn}} \cos\alpha  + (1 - e^{\mathrm{i}2\omega_{mn}})\sin\alpha ) \big)|1\rangle.
\end{split}
\end{equation}
The real output state will be
\begin{equation}\label{eq44}
\begin{split}
|\zeta_{\text{out}}\rangle_{\text{real}} = &
   \big( \frac{1}{(1+\epsilon)^2+1} \big( ((1+\epsilon)^2       e^{\mathrm{i}2(\omega_{mn}-\Delta\omega_{mn})} - 1) e^{\mathrm{i}(\varphi_{mn}-\Delta\varphi_{mn})} \cos\alpha  \\ &
                                                       + \mathrm{i}(1+\epsilon) (e^{\text{i}2(\omega_{mn}-\Delta\omega_{mn})} + 1) \sin\alpha  \big) \big) |0\rangle \\ &
 +\big( \frac{1}{(1+\epsilon)^2+1}  \big(\mathrm{i}(1+\epsilon) (e^{\text{i}2(\omega_{mn}-\Delta\omega_{mn})} + 1) 
                                                     e^{\mathrm{i}(\varphi_{mn}-\Delta\varphi_{mn})} \cos\alpha  \\&
                                                                + ((1+\epsilon)^2 - e^{\mathrm{i}2(\omega_{mn}-\Delta\omega_{mn})}) \sin\alpha \big) \big)|1\rangle.
\end{split}
\end{equation}
The quality of the block $T_{mn}$ is characterized through the fidelity, which is defined as
\begin{equation}\label{eq45}
\mathcal{F}_{T_{mn}} = \frac{1}{2\pi} \int_0^{2\pi} |{}_{\text{idea}}\langle \zeta_{\text{out}}|\zeta_{\text{out}}\rangle_{\text{real}}|^2 \text{d} \alpha.
\end{equation}
The average fidelity of $T_{mn}$ is shown in Fig. \ref{Fig4} as function of $\epsilon$ and $\Delta\omega_{mn}$.

\section{Evaluation of the feasibility of proposed qutrit ECP scheme}      \label{sec4}

\subsection{The imperfections arising from linear optics and cross-Kerr nonlinearities}

The single-partite projection measurement in the basis $\{|\phi_0\rangle,|\phi_1\rangle,|\phi_2\rangle\}$ and the cross-Kerr nonlinearities are the two key elements in our scheme.
The measurements can be implemented by using Fig. \ref{Fig2}.
Hence, the error arises from the imperfection 50:50 BSs, phase shifters, and single-photon detectors (the detector dark count) will degrade the performance of our scheme, see Fig. \ref{Fig4}.
We believe these imperfections may be mitigated in near future.

\begin{figure}[htbp]
\centering\includegraphics[width=8.5 cm]{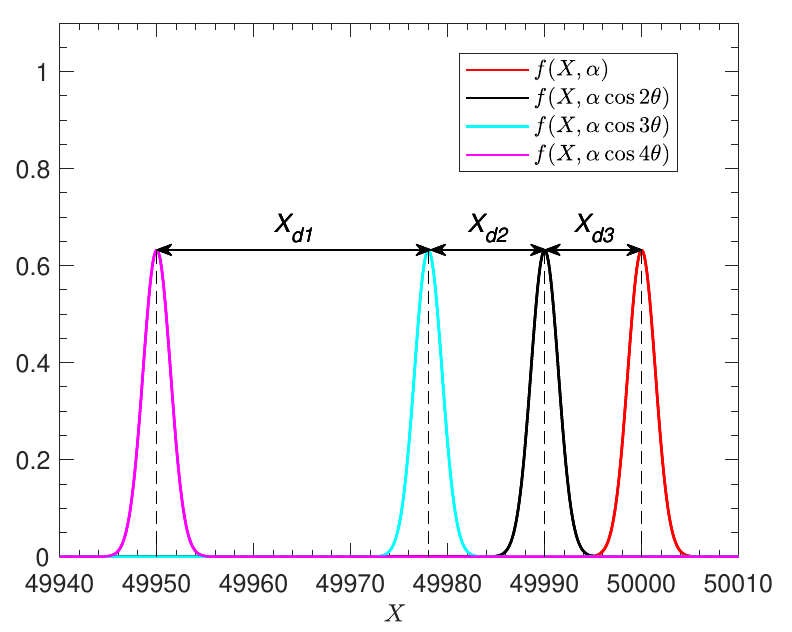}
\caption{Gaussian function curves of $f(X,\alpha)$ (red line), $f(X,\alpha\cos2\theta)$ (black line), $f(X,\alpha\cos3\theta)$ (blue line), and $f(X,\alpha\cos4\theta)$ (purple line). Here $\alpha=5000$ and $\theta=10^{-2}$ are taken.}
\label{Fig5}
\end{figure}

Now we consider the feasibility of the cross-Kerr nonlinearity.
In our ECP scheme, four possible measurement outcomes of the coherent state $|\alpha\rangle$, i.e., $0$, $2\theta$, $3\theta$, and $4\theta$ are taken into account.
The peaks of Gaussian curves $f(X,\alpha \cos (i\theta))$ $(i=0,\;2,\;3,\;4)$ at $2\alpha$, $2\alpha\cos(2\theta)$, $2\alpha\cos(3\theta)$, and $2\alpha\cos(4\theta)$.
The Gaussian function curves $f(X,\alpha)$ and $f(X,\alpha\cos(2\theta))$ exhibit partial overlap, which lead to errors between phase shifts $0$ and $2\theta$ shown in Fig. \ref{Fig5}.
Note that the influence of $f(X,\alpha\cos(3\theta))$ and $f(X,\alpha\cos(4\theta))$ can be disregarded because the Gaussian curves $f(X,\alpha\cos3\theta)$ and $f(X,\alpha\cos4\theta)$ are far away from $f(X,\alpha)$.
The $X_{d_{n}}$ given in Fig. \ref{Fig5} represents the distance between two adjacent Gaussian curves.

Let us analysis the success probability of our $X$-homodyne measurement.
Considering the photon dissipation of the coherent state, the success probability $P_{\mathrm{suc}}$ of the homodyne measurement is given by
\begin{equation}\label{46}
   P_{\mathrm{suc}}(\alpha,\theta,\gamma t) = 1-\frac{1}{2}{\mathrm{erfc}}[\frac{e^{\frac{-\gamma t}{2}\alpha(1-\cos\theta)}}{\sqrt{2}}],
\end{equation}
where $\gamma$ is the decay constant, $\mathrm{erfc}(x)$ is the Gauss complementary error function.
In our ECP scheme, we introduce two $X$-homodyne measurements.
Considering the cascade effects of the homodyne detection, the success probability of two $X$-homodyne measurements in our scheme can be calculated as $P_X=P^2_{\mathrm{suc}}$.

\begin{figure}[htbp]
\centering\includegraphics[width=10 cm]{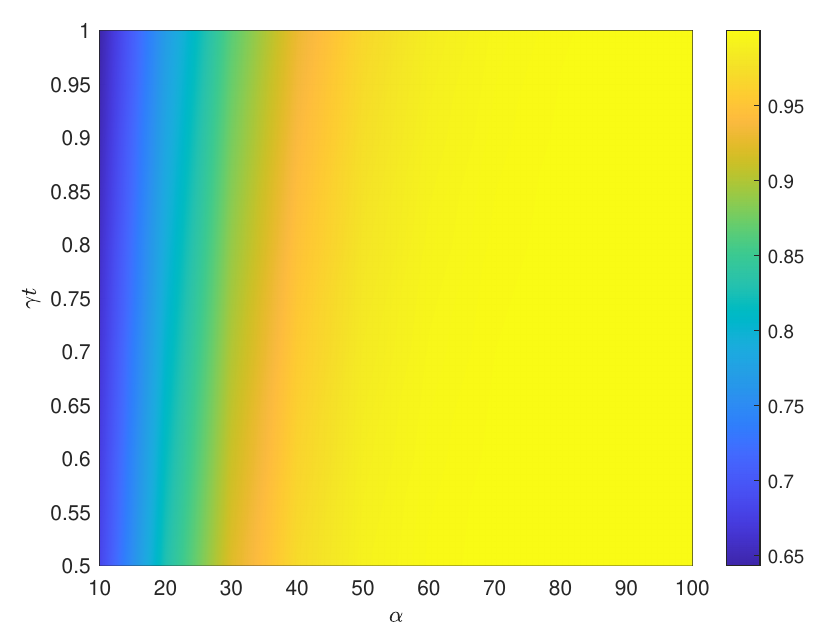}
\caption{The success probability of $X$-homodyne measurement $P_X$ as a function of $\alpha$ and $\gamma t$ in the qutrit ECP scheme.  Here $\theta=0.35$  \cite{035} is taken.}
\label{Fig6}
\end{figure}

\begin{figure}[htbp]
\centering\includegraphics[width=8.5 cm]{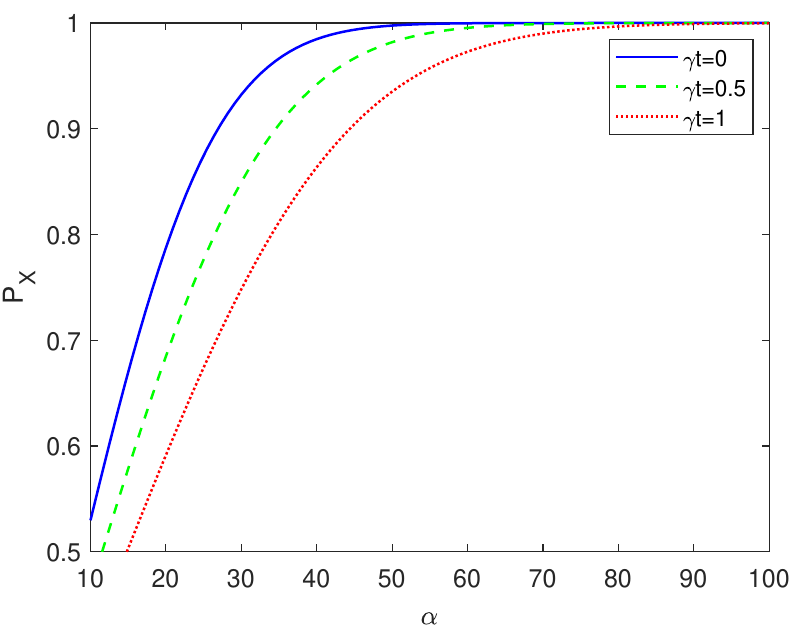}
\caption{The success probability of $X$-homodyne measurement $P_X$ as a function of $\alpha$ in the qutrit ECP scheme.  Here $\theta=0.35$  \cite{035} is taken.}
\label{Fig7}
\end{figure}

Based on Fig. \ref{Fig6} and Fig. \ref{Fig7}, we can find that the high success probability of $X$-homodyne measurement $P_X$ can be obtained with large amplitude $\alpha$ of the coherent state and lower dissipation coefficient $\gamma t$.
For example, $P_{\mathrm{suc}}(50,1)=0.9351$, $P_{\mathrm{suc}}(50,0.5)=0.9819$, $P_{\mathrm{suc}}(50,0)=0.9976$, $P_{\mathrm{suc}}(60,1)=0.9728$, $P_{\mathrm{suc}}(60,0.5)=0.9953$, $P_{\mathrm{suc}}(60,0)=0.9997$.
The above data indicates that the $X$-homodyne measurement in our ECP scheme is feasible.

\subsection{High-dimensional ECP with known parameters}

If the parameters $\alpha$, $\beta$, and $\gamma$ of the normalization less-entangled state
\begin{equation}\label{47}
   |\Phi\rangle = \alpha|0_a 0_b\rangle + \beta|1_a 1_b\rangle + \gamma|2_a 2_b\rangle,
\end{equation}
are known to Alice and Bob, $|\Phi\rangle$ can be distilled into the maximally entangled state $\frac{1}{\sqrt{3}}(|0_a 0_b\rangle + |1_a 1_b\rangle + |2_a 2_b\rangle)$ by solely employing linear optics. Here $|\alpha|>|\beta|>|\gamma|$.

\begin{figure}[htbp]
\centering\includegraphics[width=6.5 cm]{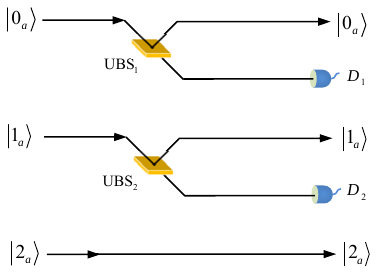}
\caption{Schematic diagram of our ECP for a two photon partially entangled qutrit state with known parameters in the spatial-mode DOF. UBS$_i$ ($i=1,2$) represent two unbalanced beam splitters with reflection coefficient $R_1=\frac{\gamma}{\alpha}, R_2=\frac{\gamma}{\beta}$, respectively.}
\label{Fig8}
\end{figure}

As shown in Fig. \ref{Fig8}, Alice first arranges an unbalanced beam splitter UBS$_1$ with a reflectivity of $r_1=\frac{\gamma}{\alpha}$ on the spatial mode $|0_a\rangle$ to balance the parameter $\alpha$. That is UBS$_1$ converts $|\Phi\rangle$ to
\begin{equation}\label{48}
   |\Phi_1\rangle =  \gamma(|0_a 0_b\rangle + |2_a 2_b\rangle) + \beta|1_a 1_b\rangle  + \sqrt{\alpha^2-\gamma^2} |0_{a'} 0_b\rangle.
\end{equation}

Subsequently, Alice arranges UBS$_2$ with a reflectivity of $r_2=\frac{\gamma}{\beta}$ on the spatial mode $|1_a\rangle$, which can convert $|\Phi_1\rangle$ to
\begin{equation}\label{49}
 \begin{split}
   |\Phi_2\rangle   & = \gamma(|0_a 0_b\rangle + |1_a 1_b\rangle + |2_a 2_b\rangle) 
                         + \sqrt{\alpha^2-\beta^2}|0_{a'} 0_b\rangle + \sqrt{\beta^2-\gamma^2}|1_{a'} 1_b\rangle.
  \end{split}
\end{equation}

Lastly, the photons in the spatial state $|0_{a'}\rangle$ and $|1_{a'}\rangle$ are detected by the single-photon detectors $D_1$ and $D_2$, respectively.
Based on Eq. \eqref{49}, one can see that if none of the detectors are fired, then $|\Phi_2\rangle$ will collapse to the desired state
\begin{equation}\label{50}
   |\Phi'\rangle = \frac{1}{\sqrt{3}}(|0_a 0_b\rangle + |1_a 1_b\rangle + |2_a 2_b\rangle).
\end{equation}
with a success probability of $\frac{|\gamma|^2}{3}$.

\medskip

\section{Conclusion}      \label{sec5}

We have proposed a scheme to implement an ECP for a two-photon system in a high-dimensional unknown two-qutrit partially entangled state.
We use the cross-Kerr nonlinearities and single-partite projection measurements to postselect two photons with two-qutrit maximally entangled state.
Our ECP is introduced in high-dimensional three-level systems, while the previous works are focused on two-level qubit systems.
In the proposed ECP, all operations are performed only at Bob's site. 
Note that the undesired less-level partially entangled states in subsystems are not useless, and they are the fascinating resources for qubit ECP, 
circuit cutting, deterministic hierarchical sharing, controller-independent quantum bidirectional communication, and quantum key distribution.
Moreover, single-qutrit projection measurements, which are the key component of our qutrit ECP with unknown parameters, are implemented by solely using linear optics. 
A linear optical architecture for implementing high-dimensional ECP with known parameters is also proposed.
The evaluation of the feasibility of our qutrit ECP shows that the proposed ECP has the potential for certain high-dimensional QIP tasks.
In near future, we will study the high-dimensional EPP protocol, which is used to increase the entanglement of quantum systems in a mixed state.

\medskip

\section*{Funding} \par

This work is supported by the National Natural Science Foundation of China under Grant No. 62371038, the Fundamental Research Funds
for the Central Universities under Grant No. FRF-TP-19-011A3, the Tian-jin Natural Science Foundation under Grant No. 23JCQNJC00560, and the Beijing Natural Science Foundation Grant No. 4252006.

\section*{Disclosures} \par
The authors declare no conflicts of interest.

\section*{Data availability} \par

The data supporting the conclusions of this study are available from the corresponding author upon reasonable request.

\smallskip

\appendix

\section{Derivation of the Success Probability for the X-Homodyne Measurement} 

This appendix provides a detailed derivation of the success probability for the $X$-homodyne measurement, accounting for photon dissipation in the coherent state. 
The derivation is based on the evolution and measurement of Gaussian states in quantum phase space.
The protocol begins with a coherent state $| \alpha \rangle$. 
Its evolution under a photon dissipation channel, characterized by a decay constant $\gamma$, is considered. 
After time $t$, the amplitude of the initial coherent state decays and vacuum noise is introduced. 
This evolution is precisely described in phase space: the initial coherent state's Wigner function, a Gaussian peak located at $\alpha$ evolves into a broader Gaussian peak with a displaced mean \cite{book}.

The probability distribution $p(x)$ for the evolved state's $X$-quadrature remains Gaussian, with its mean $\mu_0$ and variance $\sigma^2$ given by:
\begin{eqnarray}                         \label{51}
  \begin{split}
    \mu_0 = \sqrt{2} \, \text{Re}(\alpha e^{-\gamma t / 2}), \quad
    \sigma^2 = \frac{1}{2}(2 - e^{-\gamma t}).
 \end{split}
\end{eqnarray}
The variance term $\frac{1}{2}(2 - e^{-\gamma t})$ accounts for the increased noise due to energy decay.

A displacement operation $D(\beta)$ is then applied to the system, where the displacement magnitude is $\beta = \alpha(1 - \cos\theta)$. 
This specific form originates from the quantum interference setup within our protocol. The displacement shifts the entire phase-space probability distribution. 
Consequently, the mean value of the $X$-quadrature after displacement becomes
\begin{eqnarray}                         \label{52}
  \begin{split}
    \mu &= \sqrt{2} \, \text{Re}(\alpha e^{-\gamma t / 2} + \beta).
 \end{split}
\end{eqnarray}

Assuming $\alpha$ is real for simplicity and substituting the expression for $\beta$, the final mean is:
\begin{eqnarray}                         \label{53}
  \begin{split}
    \mu &= \sqrt{2} \, \alpha ( e^{-\gamma t / 2} + 1 - \cos\theta ).
 \end{split}
\end{eqnarray}
The displacement operation does not alter the variance, which remains $\sigma^2 = \frac{1}{2}(2 - e^{-\gamma t})$.

An $X$-homodyne measurement is performed on the displaced state. The measurement outcome $x$ follows a Gaussian distribution:
\begin{eqnarray}                         \label{54}
  \begin{split}
    p(x) = \frac{1}{\sqrt{2\pi\sigma^2}} \exp\left[ -\frac{(x - \mu)^2}{2\sigma^2} \right].
 \end{split}
\end{eqnarray}
In our protocol, a successful event is defined by the measurement outcome $x>0$. 
Therefore, the success probability $P_{\text{suc}}$ is obtained by integrating this probability density function over the decision region from zero to positive infinity
\begin{eqnarray}                         \label{55}
  \begin{split}
    P_{\text{suc}} = \int_{0}^{\infty} p(x)  \text{d}x = \int_{0}^{\infty} \frac{1}{\sqrt{2\pi\sigma^2}} \exp\left[ -\frac{(x - \mu)^2}{2\sigma^2} \right] \text{d}x.
 \end{split}
\end{eqnarray}

This integral represents a standard Gaussian tail probability, which can be expressed using the complementary error function $\text{erfc}(z)$
\begin{eqnarray}                         \label{56}
  \begin{split}
    P_{\text{suc}} = \frac{1}{2} \, \text{erfc}\left( -\frac{\mu}{4\sigma^2} \right).
 \end{split}
\end{eqnarray}
Substituting the expressions for $\mu$ and $\sigma^2$ yields
\begin{eqnarray}                         \label{57}
  \begin{split}
    P_{\text{suc}} = \frac{1}{2} \, \text{erfc}\left( -\frac{ \sqrt{2} \, \alpha ( e^{-\gamma t / 2} + 1 - \cos\theta ) }{2(2 - e^{-\gamma t})} \right).
 \end{split}
\end{eqnarray}

Using the property of the complementary error function $\text{erfc}(-z) = 2 - \text{erfc}(z)$, we transform the above result:
\begin{eqnarray}                         \label{58}
  \begin{split}
    P_{\text{suc}} &= \frac{1}{2} \left[ 2 - \text{erfc}\left( \frac{ \alpha ( e^{-\gamma t / 2} + 1 - \cos\theta ) }{ \sqrt{2}(2 - e^{-\gamma t})} \right) \right]\\
    &= 1 - \frac{1}{2} \, \text{erfc}\left( \frac{ \alpha ( e^{-\gamma t / 2} + 1 - \cos\theta ) }{ \sqrt{2}(2 - e^{-\gamma t})} \right).
 \end{split}
\end{eqnarray}

Under the low-dissipation approximation (i.e., $\gamma t \ll 1$), we have $e^{-\gamma t / 2} \approx 1$ and $2 - e^{-\gamma t} \approx 1$. Substituting these approximations into the equation above yields the simplified expression used in the main text
\begin{eqnarray}                         \label{59}
  \begin{split}
    P_{\text{suc}}(\alpha, \theta, \gamma t) = 1 - \frac{1}{2} \text{erfc} \left[ \frac{e^{-\frac{\gamma t}{2}} \alpha (1 - \cos \theta)} {\sqrt{2}} \right].
 \end{split}
\end{eqnarray}

\section{Table for qutrit ECP protocol} \par

\begin{table}[htbp]
 \caption{The correspondences between single-partite measurements, output state, and classical single-qutrit operations to obtain the desired maximally entangled state $|\Psi_{4_0}\rangle=\frac{1}{\sqrt{3}}(|0_a 0_b\rangle + |1_a 1_b\rangle + |2_a 2_b\rangle)$.
         Here $|\Omega_{0}\rangle = \frac{1}{\sqrt{3}}(|0_a 0_b\rangle + |1_a 1_b\rangle + |2_a 2_b\rangle)$,
               $|\Omega_{1}\rangle = \frac{1}{\sqrt{3}}(|0_a 0_b\rangle + e^{\frac{\mathrm{i}4\pi}{3}} |1_a 1_b\rangle + e^{\frac{\mathrm{i}2\pi}{3}} |2_a 2_b\rangle)$,
           and $|\Omega_{2}\rangle = \frac{1}{\sqrt{3}}(|0_a 0_b\rangle + e^{\frac{\mathrm{i}2\pi}{3}} |1_a 1_b\rangle + e^{\frac{\mathrm{i}4\pi}{3}} |2_a 2_b\rangle)$.} \label{Table2}
  \centering
\begin{tabular}{ccc}
 \hline\hline
  Single-photon measurements   &   Output state   &    Unitary operation   \\
 \hline
 $(|\phi_{0}\rangle_c,|\phi_{0}\rangle_e,|\phi_{0}\rangle_d,|\phi_{0}\rangle_f)$,  $(|\phi_{1}\rangle_c,|\phi_{1}\rangle_e,|\phi_{0}\rangle_d,|\phi_{1}\rangle_f)$      &                                                                                                     &    \\

  $(|\phi_{0}\rangle_c,|\phi_{0}\rangle_e,|\phi_{2}\rangle_d,|\phi_{1}\rangle_f)$, $(|\phi_{1}\rangle_c,|\phi_{2}\rangle_e,|\phi_{2}\rangle_d,|\phi_{1}\rangle_f)$       &                                                                                                                                  &   \\

  $(|\phi_{0}\rangle_c,|\phi_{0}\rangle_e,|\phi_{1}\rangle_d,|\phi_{2}\rangle_f)$, $(|\phi_{1}\rangle_c,|\phi_{2}\rangle_e,|\phi_{1}\rangle_d,|\phi_{2}\rangle_f)$       &                                                                                                                                  &   \\

  $(|\phi_{0}\rangle_c,|\phi_{1}\rangle_e,|\phi_{0}\rangle_d,|\phi_{2}\rangle_f)$, $(|\phi_{1}\rangle_c,|\phi_{2}\rangle_e,|\phi_{0}\rangle_d,|\phi_{0}\rangle_f)$       &                                                                                                                                  &   \\

  $(|\phi_{0}\rangle_c,|\phi_{1}\rangle_e,|\phi_{2}\rangle_d,|\phi_{0}\rangle_f)$, $(|\phi_{2}\rangle_c,|\phi_{0}\rangle_e,|\phi_{1}\rangle_d,|\phi_{0}\rangle_f)$       &                                                                                                                                  &   \\

  $(|\phi_{0}\rangle_c,|\phi_{1}\rangle_e,|\phi_{1}\rangle_d,|\phi_{1}\rangle_f)$, $(|\phi_{2}\rangle_c,|\phi_{0}\rangle_e,|\phi_{0}\rangle_d,|\phi_{1}\rangle_f)$       &                                                                                                                                 &    \\

  $(|\phi_{0}\rangle_c,|\phi_{2}\rangle_e,|\phi_{0}\rangle_d,|\phi_{1}\rangle_f)$, $(|\phi_{2}\rangle_c,|\phi_{0}\rangle_e,|\phi_{2}\rangle_d,|\phi_{2}\rangle_f)$       &  $|\Omega_{0}\rangle$                                                                                                                &  $I$  \\

  $(|\phi_{0}\rangle_c,|\phi_{2}\rangle_e,|\phi_{2}\rangle_d,|\phi_{2}\rangle_f)$, $(|\phi_{2}\rangle_c,|\phi_{1}\rangle_e,|\phi_{1}\rangle_d,|\phi_{2}\rangle_f)$       &                                                                                                                                  &   \\

  $(|\phi_{0}\rangle_c,|\phi_{2}\rangle_e,|\phi_{1}\rangle_d,|\phi_{0}\rangle_f)$, $(|\phi_{2}\rangle_c,|\phi_{1}\rangle_e,|\phi_{0}\rangle_d,|\phi_{0}\rangle_f)$       &  $ $                                                                                                                                 &   \\

  $(|\phi_{1}\rangle_c,|\phi_{0}\rangle_e,|\phi_{2}\rangle_d,|\phi_{0}\rangle_f)$, $(|\phi_{2}\rangle_c,|\phi_{1}\rangle_e,|\phi_{2}\rangle_d,|\phi_{1}\rangle_f)$       &  $ $                                                                                                                                 &   \\

  $(|\phi_{1}\rangle_c,|\phi_{0}\rangle_e,|\phi_{1}\rangle_d,|\phi_{1}\rangle_f)$, $(|\phi_{2}\rangle_c,|\phi_{2}\rangle_e,|\phi_{1}\rangle_d,|\phi_{1}\rangle_f)$       &  $ $                                                                                                                                 &   \\

  $(|\phi_{1}\rangle_c,|\phi_{0}\rangle_e,|\phi_{0}\rangle_d,|\phi_{2}\rangle_f)$, $(|\phi_{2}\rangle_c,|\phi_{2}\rangle_e,|\phi_{0}\rangle_d,|\phi_{2}\rangle_f)$       &  $ $                                                                                                                                 &    \\

  $(|\phi_{1}\rangle_c,|\phi_{1}\rangle_e,|\phi_{2}\rangle_d,|\phi_{2}\rangle_f)$, $(|\phi_{2}\rangle_c,|\phi_{2}\rangle_e,|\phi_{2}\rangle_d,|\phi_{0}\rangle_f)$       &  $ $                                                                                                                                 &   \\

  $(|\phi_{1}\rangle_c,|\phi_{1}\rangle_e,|\phi_{1}\rangle_d,|\phi_{0}\rangle_f)$                                                                                        &  $ $                                                                                                                                 &    \\
\hline
  $(|\phi_{0}\rangle_c,|\phi_{0}\rangle_e,|\phi_{1}\rangle_d,|\phi_{0}\rangle_f)$, $(|\phi_{2}\rangle_c,|\phi_{1}\rangle_e,|\phi_{2}\rangle_d,|\phi_{1}\rangle_f)$      &  $ $                                                                                                                                  &    \\

  $(|\phi_{1}\rangle_c,|\phi_{0}\rangle_e,|\phi_{0}\rangle_d,|\phi_{0}\rangle_f)$, $(|\phi_{0}\rangle_c,|\phi_{2}\rangle_e,|\phi_{1}\rangle_d,|\phi_{2}\rangle_f)$      &  $ $                                                                                                                                  &   \\

  $(|\phi_{2}\rangle_c,|\phi_{0}\rangle_e,|\phi_{2}\rangle_d,|\phi_{0}\rangle_f)$, $(|\phi_{1}\rangle_c,|\phi_{2}\rangle_e,|\phi_{0}\rangle_d,|\phi_{2}\rangle_f)$      &  $|\Omega_{1}\rangle$                                                                                                                 &  $U_1$ \\

  $(|\phi_{0}\rangle_c,|\phi_{1}\rangle_e,|\phi_{1}\rangle_d,|\phi_{1}\rangle_f)$, $(|\phi_{2}\rangle_c,|\phi_{2}\rangle_e,|\phi_{2}\rangle_d,|\phi_{2}\rangle_f)$      &  $ $                                                                                                                                  &    \\

  $(|\phi_{1}\rangle_c,|\phi_{1}\rangle_e,|\phi_{0}\rangle_d,|\phi_{1}\rangle_f)$                                                                                       &  $ $                                                                                                                                  &    \\
\hline
  $(|\phi_{0}\rangle_c,|\phi_{0}\rangle_e,|\phi_{2}\rangle_d,|\phi_{0}\rangle_f)$, $(|\phi_{2}\rangle_c,|\phi_{1}\rangle_e,|\phi_{0}\rangle_d,|\phi_{1}\rangle_f)$    &  $ $                                                                                                                                    &    \\

  $(|\phi_{1}\rangle_c,|\phi_{0}\rangle_e,|\phi_{1}\rangle_d,|\phi_{0}\rangle_f)$, $(|\phi_{0}\rangle_c,|\phi_{2}\rangle_e,|\phi_{2}\rangle_d,|\phi_{2}\rangle_f)$    &  $ $                                                                                                                                    &    \\

  $(|\phi_{2}\rangle_c,|\phi_{0}\rangle_e,|\phi_{0}\rangle_d,|\phi_{0}\rangle_f)$, $(|\phi_{1}\rangle_c,|\phi_{2}\rangle_e,|\phi_{1}\rangle_d,|\phi_{2}\rangle_f)$    &  $|\Omega_{2}\rangle$                                                                                                                   &  $U_2$ \\

  $(|\phi_{0}\rangle_c,|\phi_{1}\rangle_e,|\phi_{2}\rangle_d,|\phi_{1}\rangle_f)$, $(|\phi_{2}\rangle_c,|\phi_{2}\rangle_e,|\phi_{0}\rangle_d,|\phi_{2}\rangle_f)$    &  $ $                                                                                                                                    &  $ $ \\

  $(|\phi_{1}\rangle_c,|\phi_{1}\rangle_e,|\phi_{1}\rangle_d,|\phi_{1}\rangle_f)$                                                                                     &  $ $                                                                                                                                    &  $ $  \\
\hline\hline
\end{tabular}
\end{table}

\smallskip

\begin{table}[htbp]
\caption{The correspondences between single-partite measurements, output state, and classical single-qutrit operations to obtain the desired maximally entangled state $|\Psi_{4_0}\rangle=\frac{1}{\sqrt{3}}(|0_a 0_b\rangle + |1_a 1_b\rangle + |2_a 2_b\rangle)$.
         Here $|\Omega_{3}\rangle = \frac{1}{\sqrt{3}}(-|0_a 0_b\rangle + e^{\frac{\mathrm{i}5\pi}{3}}|1_a 1_b\rangle + e^{\frac{\mathrm{i}\pi}{3}}|2_a 2_b\rangle)$,
               $|\Omega_{4}\rangle = \frac{1}{\sqrt{3}}(-|0_a 0_b\rangle + e^{\frac{\mathrm{i}\pi}{3}}|1_a 1_b\rangle  + e^{\frac{\mathrm{i}5\pi}{3}}|2_a 2_b\rangle)$.}
  \label{Table3}
  \centering
\begin{tabular}{ccc}
\hline\hline
Single-photon measurements   &   Output state   &    Unitary operation\\
\hline
  $(|\phi_{0}\rangle_c,|\phi_{0}\rangle_e,|\phi_{1}\rangle_d,|\phi_{0}\rangle_f)$, $(|\phi_{1}\rangle_c,|\phi_{1}\rangle_e,|\phi_{2}\rangle_d,|\phi_{0}\rangle_f)$       &  $ $                                                                                                                                 &  $ $ \\

  $(|\phi_{0}\rangle_c,|\phi_{0}\rangle_e,|\phi_{0}\rangle_d,|\phi_{1}\rangle_f)$, $(|\phi_{1}\rangle_c,|\phi_{2}\rangle_e,|\phi_{0}\rangle_d,|\phi_{1}\rangle_f)$       &  $ $                                                                                                                                 &  $ $ \\

  $(|\phi_{0}\rangle_c,|\phi_{1}\rangle_e,|\phi_{1}\rangle_d,|\phi_{2}\rangle_f)$, $(|\phi_{1}\rangle_c,|\phi_{2}\rangle_e,|\phi_{2}\rangle_d,|\phi_{2}\rangle_f)$       &  $ $                                                                                                                                 &  $ $ \\

  $(|\phi_{0}\rangle_c,|\phi_{1}\rangle_e,|\phi_{0}\rangle_d,|\phi_{0}\rangle_f)$, $(|\phi_{2}\rangle_c,|\phi_{0}\rangle_e,|\phi_{2}\rangle_d,|\phi_{0}\rangle_f)$       &  $ $                                                                                                                                 &  $ $ \\

  $(|\phi_{0}\rangle_c,|\phi_{2}\rangle_e,|\phi_{1}\rangle_d,|\phi_{1}\rangle_f)$, $(|\phi_{2}\rangle_c,|\phi_{0}\rangle_e,|\phi_{1}\rangle_d,|\phi_{1}\rangle_f)$       &  $|\Omega_{3}\rangle$                                                                                                                &  $U_3$ \\

  $(|\phi_{0}\rangle_c,|\phi_{2}\rangle_e,|\phi_{0}\rangle_d,|\phi_{2}\rangle_f)$, $(|\phi_{2}\rangle_c,|\phi_{1}\rangle_e,|\phi_{2}\rangle_d,|\phi_{2}\rangle_f)$       &  $ $                                                                                                                                 &  $ $ \\

  $(|\phi_{1}\rangle_c,|\phi_{0}\rangle_e,|\phi_{0}\rangle_d,|\phi_{0}\rangle_f)$, $(|\phi_{2}\rangle_c,|\phi_{1}\rangle_e,|\phi_{1}\rangle_d,|\phi_{0}\rangle_f)$       &  $ $                                                                                                                                 &  $ $ \\

  $(|\phi_{1}\rangle_c,|\phi_{0}\rangle_e,|\phi_{2}\rangle_d,|\phi_{1}\rangle_f)$, $(|\phi_{2}\rangle_c,|\phi_{2}\rangle_e,|\phi_{2}\rangle_d,|\phi_{1}\rangle_f)$       &  $ $                                                                                                                                 &  $ $ \\

  $(|\phi_{1}\rangle_c,|\phi_{1}\rangle_e,|\phi_{0}\rangle_d,|\phi_{2}\rangle_f)$, $(|\phi_{2}\rangle_c,|\phi_{2}\rangle_e,|\phi_{1}\rangle_d,|\phi_{2}\rangle_f)$       &  $ $                                                                                                                                 &  $ $ \\
\hline
  $(|\phi_{0}\rangle_c,|\phi_{0}\rangle_e,|\phi_{2}\rangle_d,|\phi_{0}\rangle_f)$, $(|\phi_{1}\rangle_c,|\phi_{1}\rangle_e,|\phi_{2}\rangle_d,|\phi_{1}\rangle_f)$       &  $ $                                                                                                                                 &  $ $ \\

  $(|\phi_{0}\rangle_c,|\phi_{0}\rangle_e,|\phi_{0}\rangle_d,|\phi_{2}\rangle_f)$, $(|\phi_{1}\rangle_c,|\phi_{2}\rangle_e,|\phi_{1}\rangle_d,|\phi_{1}\rangle_f)$       &  $ $                                                                                                                                 &  $ $ \\

  $(|\phi_{0}\rangle_c,|\phi_{1}\rangle_e,|\phi_{2}\rangle_d,|\phi_{2}\rangle_f)$, $(|\phi_{1}\rangle_c,|\phi_{2}\rangle_e,|\phi_{2}\rangle_d,|\phi_{0}\rangle_f)$       &  $ $                                                                                                                                 &  $ $ \\

  $(|\phi_{0}\rangle_c,|\phi_{1}\rangle_e,|\phi_{0}\rangle_d,|\phi_{1}\rangle_f)$, $(|\phi_{2}\rangle_c,|\phi_{0}\rangle_e,|\phi_{0}\rangle_d,|\phi_{0}\rangle_f)$       &  $ $                                                                                                                                 &  $ $ \\

  $(|\phi_{0}\rangle_c,|\phi_{2}\rangle_e,|\phi_{2}\rangle_d,|\phi_{1}\rangle_f)$, $(|\phi_{2}\rangle_c,|\phi_{0}\rangle_e,|\phi_{1}\rangle_d,|\phi_{2}\rangle_f)$       &  $|\Omega_{4}\rangle$                                                                                                                &  $U_4$ \\

  $(|\phi_{0}\rangle_c,|\phi_{2}\rangle_e,|\phi_{0}\rangle_d,|\phi_{0}\rangle_f)$, $(|\phi_{2}\rangle_c,|\phi_{1}\rangle_e,|\phi_{0}\rangle_d,|\phi_{2}\rangle_f)$       &  $ $                                                                                                                                 &  $ $ \\

  $(|\phi_{1}\rangle_c,|\phi_{0}\rangle_e,|\phi_{1}\rangle_d,|\phi_{0}\rangle_f)$, $(|\phi_{2}\rangle_c,|\phi_{1}\rangle_e,|\phi_{1}\rangle_d,|\phi_{1}\rangle_f)$       &  $ $                                                                                                                                 &  $ $ \\

  $(|\phi_{1}\rangle_c,|\phi_{0}\rangle_e,|\phi_{2}\rangle_d,|\phi_{2}\rangle_f)$, $(|\phi_{2}\rangle_c,|\phi_{2}\rangle_e,|\phi_{0}\rangle_d,|\phi_{1}\rangle_f)$       &  $ $                                                                                                                                 &  $ $ \\

  $(|\phi_{1}\rangle_c,|\phi_{1}\rangle_e,|\phi_{1}\rangle_d,|\phi_{2}\rangle_f)$, $(|\phi_{2}\rangle_c,|\phi_{2}\rangle_e,|\phi_{1}\rangle_d,|\phi_{0}\rangle_f)$       &  $ $                                                                                                                                 &  $ $ \\
\hline\hline
\end{tabular}
\end{table}

\begin{table}[htbp]
\caption{The relations between the coefficients, the $|X\rangle \langle X|$, the single-partite projection measurements, and the concentrated  partially entangled qubit states
          $|\Omega_{5}\rangle = \frac{2}{\sqrt{5}}|0_a 0_b\rangle + \frac{1}{\sqrt{5}}|1_a 1_b\rangle$,
          $|\Omega_{6}\rangle = \frac{1}{\sqrt{5}}|0_a 0_b\rangle + \frac{2}{\sqrt{5}}|1_a 1_b\rangle$,
          $|\Omega_{7}\rangle = \frac{2}{\sqrt{5}}|0_a 0_b\rangle + \frac{1}{\sqrt{5}}|2_a 2_b\rangle$,
          $|\Omega_{8}\rangle = \frac{1}{\sqrt{5}}|0_a 0_b\rangle + \frac{2}{\sqrt{5}}|2_a 2_b\rangle$,
          $|\Omega_{9}\rangle = \frac{2}{\sqrt{5}}|0_a 0_b\rangle + \frac{1}{\sqrt{5}}|2_a 2_b\rangle$, and
          $|\Omega_{10}\rangle = \frac{1}{\sqrt{5}}|0_a 0_b\rangle + \frac{2}{\sqrt{5}}|2_a 2_b\rangle$.}
  \label{Table4}
  \centering
\begin{tabular}{cccc}
\hline\hline
Coefficient  &  $|X\rangle \langle X|$  &  Single-partite measurements  &  Result\\
\hline
                         &                                                                                      &  $(|\phi_{0}\rangle_c,|\phi_{0}\rangle_e,|\phi_{0}\rangle_d,|\phi_{0}\rangle_f)$, $(|\phi_{2}\rangle_c,|\phi_{0}\rangle_e,|\phi_{1}\rangle_d,|\phi_{0}\rangle_f)$,       &   \\

                         &                                                                                      &  $(|\phi_{1}\rangle_c,|\phi_{0}\rangle_e,|\phi_{2}\rangle_d,|\phi_{0}\rangle_f)$, $(|\phi_{0}\rangle_c,|\phi_{2}\rangle_e,|\phi_{0}\rangle_d,|\phi_{1}\rangle_f)$,       &   \\

      $\alpha^2\beta$    &  $|\alpha e^{-2\mathrm{i}\theta}\rangle |\alpha'e^{\mathrm{i}\theta^{'}}\rangle$     &
                            $(|\phi_{2}\rangle_c,|\phi_{2}\rangle_e,|\phi_{1}\rangle_d,|\phi_{1}\rangle_f)$, 
                            $(|\phi_{1}\rangle_c,|\phi_{2}\rangle_e,|\phi_{2}\rangle_d,|\phi_{1}\rangle_f)$,    &
      $|\Omega_{5}\rangle$ \\

                         &                                                                                      &  $(|\phi_{0}\rangle_c,|\phi_{1}\rangle_e,|\phi_{0}\rangle_d,|\phi_{2}\rangle_f)$, $(|\phi_{2}\rangle_c,|\phi_{1}\rangle_e,|\phi_{1}\rangle_d,|\phi_{2}\rangle_f)$,       &    \\

                         &                                                                                      &  $(|\phi_{1}\rangle_c,|\phi_{1}\rangle_e,|\phi_{2}\rangle_d,|\phi_{2}\rangle_f)$  &   \\
\hline

                         &                                                                                       &  $(|\phi_{0}\rangle_c,|\phi_{0}\rangle_e,|\phi_{0}\rangle_d,|\phi_{0}\rangle_f)$, $(|\phi_{2}\rangle_c,|\phi_{0}\rangle_e,|\phi_{1}\rangle_d,|\phi_{0}\rangle_f)$,        &    \\

                         &                                                                                       &  $(|\phi_{1}\rangle_c,|\phi_{0}\rangle_e,|\phi_{2}\rangle_d,|\phi_{0}\rangle_f)$, $(|\phi_{0}\rangle_c,|\phi_{2}\rangle_e,|\phi_{0}\rangle_d,|\phi_{1}\rangle_f)$,        &    \\

      $\alpha\beta^2$    &  $|\alpha e^{-\mathrm{i}\theta}\rangle |\alpha'e^{2\mathrm{i}\theta^{'}}\rangle$      &
                            $(|\phi_{2}\rangle_c,|\phi_{2}\rangle_e,|\phi_{1}\rangle_d,|\phi_{1}\rangle_f)$, 
                            $(|\phi_{1}\rangle_c,|\phi_{2}\rangle_e,|\phi_{2}\rangle_d,|\phi_{1}\rangle_f)$,     &
      $|\Omega_{6}\rangle$ \\

                         &                                                                                       &  $(|\phi_{0}\rangle_c,|\phi_{1}\rangle_e,|\phi_{0}\rangle_d,|\phi_{2}\rangle_f)$, $(|\phi_{2}\rangle_c,|\phi_{1}\rangle_e,|\phi_{1}\rangle_d,|\phi_{2}\rangle_f)$,        &   \\

                         &                                                                                       &  $(|\phi_{1}\rangle_c,|\phi_{1}\rangle_e,|\phi_{2}\rangle_d,|\phi_{2}\rangle_f)$  &  \\
\hline

                         &                                                                                       &  $(|\phi_{0}\rangle_c,|\phi_{0}\rangle_e,|\phi_{0}\rangle_d,|\phi_{0}\rangle_f)$, $(|\phi_{2}\rangle_c,|\phi_{0}\rangle_e,|\phi_{1}\rangle_d,|\phi_{0}\rangle_f)$,        &    \\

                         &                                                                                       &  $(|\phi_{1}\rangle_c,|\phi_{0}\rangle_e,|\phi_{2}\rangle_d,|\phi_{0}\rangle_f)$, $(|\phi_{0}\rangle_c,|\phi_{2}\rangle_e,|\phi_{0}\rangle_d,|\phi_{1}\rangle_f)$,        &    \\

      $\alpha^2\gamma$   &  $|\alpha\rangle |\alpha'e^{2\mathrm{i}\theta^{'}}\rangle$                            &
                            $(|\phi_{2}\rangle_c,|\phi_{2}\rangle_e,|\phi_{1}\rangle_d,|\phi_{1}\rangle_f)$, 
                            $(|\phi_{1}\rangle_c,|\phi_{2}\rangle_e,|\phi_{2}\rangle_d,|\phi_{1}\rangle_f)$,     &
      $|\Omega_{7}\rangle$ \\

                         &                                                                                       &  $(|\phi_{0}\rangle_c,|\phi_{1}\rangle_e,|\phi_{0}\rangle_d,|\phi_{2}\rangle_f)$, $(|\phi_{2}\rangle_c,|\phi_{1}\rangle_e,|\phi_{1}\rangle_d,|\phi_{2}\rangle_f)$,        &   \\

                         &                                                                                       &  $(|\phi_{1}\rangle_c,|\phi_{1}\rangle_e,|\phi_{2}\rangle_d,|\phi_{2}\rangle_f)$  &   \\
\hline

                         &                                                                                       &  $(|\phi_{0}\rangle_c,|\phi_{0}\rangle_e,|\phi_{0}\rangle_d,|\phi_{0}\rangle_f)$, $(|\phi_{2}\rangle_c,|\phi_{0}\rangle_e,|\phi_{1}\rangle_d,|\phi_{0}\rangle_f)$,        &    \\

                         &                                                                                       &  $(|\phi_{1}\rangle_c,|\phi_{0}\rangle_e,|\phi_{2}\rangle_d,|\phi_{0}\rangle_f)$, $(|\phi_{0}\rangle_c,|\phi_{2}\rangle_e,|\phi_{0}\rangle_d,|\phi_{1}\rangle_f)$,        &    \\

       $\alpha\gamma^2$   &  $|\alpha e^{3\mathrm{i}\theta}\rangle |\alpha'e^{4\mathrm{i}\theta^{'}}\rangle$     &
                             $(|\phi_{2}\rangle_c,|\phi_{2}\rangle_e,|\phi_{1}\rangle_d,|\phi_{1}\rangle_f)$, 
                             $(|\phi_{1}\rangle_c,|\phi_{2}\rangle_e,|\phi_{2}\rangle_d,|\phi_{1}\rangle_f)$,    &
       $|\Omega_{8}\rangle$ \\

                         &                                                                                       &  $(|\phi_{0}\rangle_c,|\phi_{1}\rangle_e,|\phi_{0}\rangle_d,|\phi_{2}\rangle_f)$, $(|\phi_{2}\rangle_c,|\phi_{1}\rangle_e,|\phi_{1}\rangle_d,|\phi_{2}\rangle_f)$,        &   \\

                         &                                                                                       &  $(|\phi_{1}\rangle_c,|\phi_{1}\rangle_e,|\phi_{2}\rangle_d,|\phi_{2}\rangle_f)$  &    \\
\hline

                         &                                                                                      &  $(|\phi_{0}\rangle_c,|\phi_{0}\rangle_e,|\phi_{0}\rangle_d,|\phi_{0}\rangle_f)$, $(|\phi_{2}\rangle_c,|\phi_{0}\rangle_e,|\phi_{1}\rangle_d,|\phi_{0}\rangle_f)$,       &   \\

                         &                                                                                      &  $(|\phi_{1}\rangle_c,|\phi_{0}\rangle_e,|\phi_{2}\rangle_d,|\phi_{0}\rangle_f)$, $(|\phi_{0}\rangle_c,|\phi_{2}\rangle_e,|\phi_{0}\rangle_d,|\phi_{1}\rangle_f)$,       &   \\

      $\beta^2\gamma$    &  $|\alpha e^{2\mathrm{i}\theta}\rangle |\alpha'e^{4\mathrm{i}\theta^{'}}\rangle$     &
                            $(|\phi_{2}\rangle_c,|\phi_{2}\rangle_e,|\phi_{1}\rangle_d,|\phi_{1}\rangle_f)$, 
                            $(|\phi_{1}\rangle_c,|\phi_{2}\rangle_e,|\phi_{2}\rangle_d,|\phi_{1}\rangle_f)$,    &
      $|\Omega_{9}\rangle$ \\

                         &                                                                                      &  $(|\phi_{0}\rangle_c,|\phi_{1}\rangle_e,|\phi_{0}\rangle_d,|\phi_{2}\rangle_f)$, $(|\phi_{2}\rangle_c,|\phi_{1}\rangle_e,|\phi_{1}\rangle_d,|\phi_{2}\rangle_f)$,       &   \\

                         &                                                                                      &  $(|\phi_{1}\rangle_c,|\phi_{1}\rangle_e,|\phi_{2}\rangle_d,|\phi_{2}\rangle_f)$  &  \\ 
\hline

                         &                                                                                      &  $(|\phi_{0}\rangle_c,|\phi_{0}\rangle_e,|\phi_{0}\rangle_d,|\phi_{0}\rangle_f)$, $(|\phi_{2}\rangle_c,|\phi_{0}\rangle_e,|\phi_{1}\rangle_d,|\phi_{0}\rangle_f)$,       &   \\

                         &                                                                                      &  $(|\phi_{1}\rangle_c,|\phi_{0}\rangle_e,|\phi_{2}\rangle_d,|\phi_{0}\rangle_f)$, $(|\phi_{0}\rangle_c,|\phi_{2}\rangle_e,|\phi_{0}\rangle_d,|\phi_{1}\rangle_f)$,       &    \\

      $\beta\gamma^2$    &  $|\alpha e^{4\mathrm{i}\theta}\rangle |\alpha'e^{5\mathrm{i}\theta^{'}}\rangle$     &
                            $(|\phi_{2}\rangle_c,|\phi_{2}\rangle_e,|\phi_{1}\rangle_d,|\phi_{1}\rangle_f)$, 
                            $(|\phi_{1}\rangle_c,|\phi_{2}\rangle_e,|\phi_{2}\rangle_d,|\phi_{1}\rangle_f)$,    &
      $|\Omega_{10}\rangle$ \\

                          &                                                                                     &  $(|\phi_{0}\rangle_c,|\phi_{1}\rangle_e,|\phi_{0}\rangle_d,|\phi_{2}\rangle_f)$, $(|\phi_{2}\rangle_c,|\phi_{1}\rangle_e,|\phi_{1}\rangle_d,|\phi_{2}\rangle_f)$,      &    \\

                          &                                                                                     &  $(|\phi_{1}\rangle_c,|\phi_{1}\rangle_e,|\phi_{2}\rangle_d,|\phi_{2}\rangle_f)$  &  \\
\hline\hline
\end{tabular}
\end{table}

\end{document}